\documentclass[12pt]{article}
\usepackage{amsmath,amsthm,amssymb,euscript,epsf,epsfig}
\usepackage{array}
\makeatletter
\renewcommand\section{\@startsection {section}{1}{\z@}%
                                   {-3.5ex \@plus -1ex \@minus -.2ex}
                                   {2.3ex \@plus.2ex}%
                                   {\normalfont\large\bfseries}}
\renewcommand\subsection{\@startsection{subsection}{2}{\z@}%
                                     {-3.25ex\@plus -1ex \@minus
                                     -.2ex}%
                                     {1.5ex \@plus .2ex}%
                                     {\normalfont\bfseries}}
\def\baselinestretch{1.2}
\newcommand{\be}{\begin{equation}}
\newcommand{\ee}{\end{equation}}
\newcommand{\bea}{\begin{eqnarray}\displaystyle}
\newcommand{\eea}{\end{eqnarray}}
\newcommand{\nn}{\nonumber}
\newcommand{\nnm}{\nonumber}

\makeatletter
\@addtoreset{equation}{section}
\makeatother

\def\one{{\hbox{ 1\kern-.8mm l}}}
\def\zero{{\hbox{ 0\kern-1.5mm 0}}}

\def\tC{ { \tilde C}  }

\def\mE{ \mathcal{E}}

 \def\cE{{\cal E}}

\def\DottedCircle{
\qbezier[4](0.966,-0.259)(1.04,0)(0.966,0.259)
\qbezier[4](0.966,0.259)(0.897,0.518)(0.707,0.707)
\qbezier[4](0.707,0.707)(0.518,0.897)(0.259,0.966)
\qbezier[4](0.259,0.966)(0,1.04)(-0.259,0.966)
\qbezier[4](-0.259,0.966)(-0.518,0.897)(-0.707,0.707)
\qbezier[4](-0.707,0.707)(-0.897,0.518)(-0.966,0.259)
\qbezier[4](-0.966,0.259)(-1.04,0)(-0.966,-0.259)
\qbezier[4](-0.966,-0.259)(-0.897,-0.518)(-0.707,-0.707)
\qbezier[4](-0.707,-0.707)(-0.518,-0.897)(-0.259,-0.966)
\qbezier[4](-0.259,-0.966)(0,-1.04)(0.259,-0.966)
\qbezier[4](0.259,-0.966)(0.518,-0.897)(0.707,-0.707)
\qbezier[4](0.707,-0.707)(0.897,-0.518)(0.966,-0.259)
}
\def\FullCircle{
\thicklines
\put(0,0){\circle{2}}
}
\def\Endpoint[#1]{
\ifcase#1
\put(1,0){\circle*{0.15}}
\or\put(0.866,0.5){\circle*{0.15}}
\or\put(0.5,0.866){\circle*{0.15}}
\or\put(0,1){\circle*{0.15}}
\or\put(-0.5,0.866){\circle*{0.15}}
\or\put(-0.866,0.5){\circle*{0.15}}
\or\put(-1,0){\circle*{0.15}}
\or\put(-0.866,-0.5){\circle*{0.15}}
\or\put(-0.5,-0.866){\circle*{0.15}}
\or\put(0,-1){\circle*{0.15}}
\or\put(0.5,-0.866){\circle*{0.15}}
\or\put(0.866,-0.5){\circle*{0.15}}
\fi}
\def\Arc[#1]{
\thicklines         
\ifcase#1
\qbezier[25](0.966,-0.259)(1.04,0)(0.966,0.259)
\or
\qbezier[25](0.966,0.259)(0.897,0.518)(0.707,0.707)
\or
\qbezier[25](0.707,0.707)(0.518,0.897)(0.259,0.966)
\or
\qbezier[25](0.259,0.966)(0,1.04)(-0.259,0.966)
\or
\qbezier[25](-0.259,0.966)(-0.518,0.897)(-0.707,0.707)
\or
\qbezier[25](-0.707,0.707)(-0.897,0.518)(-0.966,0.259)
\or
\qbezier[25](-0.966,0.259)(-1.04,0)(-0.966,-0.259)
\or
\qbezier[25](-0.966,-0.259)(-0.897,-0.518)(-0.707,-0.707)
\or
\qbezier[25](-0.707,-0.707)(-0.518,-0.897)(-0.259,-0.966)
\or
\qbezier[25](-0.259,-0.966)(0,-1.04)(0.259,-0.966)
\or
\qbezier[25](0.259,-0.966)(0.518,-0.897)(0.707,-0.707)
\or
\qbezier[25](0.707,-0.707)(0.897,-0.518)(0.966,-0.259)
\fi}
\def\DottedArc[#1]{
\ifcase#1
\qbezier[4](0.966,-0.259)(1.04,0)(0.966,0.259)
\or
\qbezier[4](0.966,0.259)(0.897,0.518)(0.707,0.707)
\or
\qbezier[4](0.707,0.707)(0.518,0.897)(0.259,0.966)
\or
\qbezier[4](0.259,0.966)(0,1.04)(-0.259,0.966)
\or
\qbezier[4](-0.259,0.966)(-0.518,0.897)(-0.707,0.707)
\or
\qbezier[4](-0.707,0.707)(-0.897,0.518)(-0.966,0.259)
\or
\qbezier[4](-0.966,0.259)(-1.04,0)(-0.966,-0.259)
\or
\qbezier[4](-0.966,-0.259)(-0.897,-0.518)(-0.707,-0.707)
\or
\qbezier[4](-0.707,-0.707)(-0.518,-0.897)(-0.259,-0.966)
\or
\qbezier[4](-0.259,-0.966)(0,-1.04)(0.259,-0.966)
\or
\qbezier[4](0.259,-0.966)(0.518,-0.897)(0.707,-0.707)
\or
\qbezier[4](0.707,-0.707)(0.897,-0.518)(0.966,-0.259)
\fi}
\def\Chord[#1,#2]{
\thinlines
\ifnum#1>#2\Chord[#2,#1]
\else\ifnum#1<#2
\ifcase#1
\ifcase#2
\or\qbezier(1,0)(0.516,0.138)(0.866,0.5)
\or\qbezier(1,0)(0.45,0.26)(0.5,0.866)
\or\qbezier(1,0)(0.327,0.327)(0,1)
\or\qbezier(1,0)(0.179,0.311)(-0.5,0.866)
\or\qbezier(1,0)(0.0536,0.2)(-0.866,0.5)
\or\put(1, 0){\line(-2, 0){2}}
\or\qbezier(1,0)(0.0536,-0.2)(-0.866,-0.5)
\or\qbezier(1,0)(0.179,-0.311)(-0.5,-0.866)
\or\qbezier(1,0)(0.327,-0.327)(0,-1)
\or\qbezier(1,0)(0.45,-0.26)(0.5,-0.866)
\or\qbezier(1,0)(0.516,-0.138)(0.866,-0.5)
\fi
\or\ifcase#2\or
\or\qbezier(0.866,0.5)(0.378,0.378)(0.5,0.866)
\or\qbezier(0.866,0.5)(0.26,0.45)(0,1)
\or\qbezier(0.866,0.5)(0.12,0.446)(-0.5,0.866)
\or\qbezier(0.866,0.5)(0,0.359)(-0.866,0.5)
\or\qbezier(0.866,0.5)(-0.0536,0.2)(-1,0)
\or\put(0.866, 0.5){\line(-5, -3){1.73}}
\or\qbezier(0.866,0.5)(0.146,-0.146)(-0.5,-0.866)
\or\qbezier(0.866,0.5)(0.311,-0.179)(0,-1)
\or\qbezier(0.866,0.5)(0.446,-0.12)(0.5,-0.866)
\or\qbezier(0.866,0.5)(0.52,0)(0.866,-0.5)
\fi
\or\ifcase#2\or\or
\or\qbezier(0.5,0.866)(0.138,0.516)(0,1)
\or\qbezier(0.5,0.866)(0,0.52)(-0.5,0.866)
\or\qbezier(0.5,0.866)(-0.12,0.446)(-0.866,0.5)
\or\qbezier(0.5,0.866)(-0.179,0.311)(-1,0)
\or\qbezier(0.5,0.866)(-0.146,0.146)(-0.866,-0.5)
\or\put(0.5, 0.866){\line(-3, -5){1}}
\or\qbezier(0.5,0.866)(0.2,-0.0536)(0,-1)
\or\qbezier(0.5,0.866)(0.359,0)(0.5,-0.866)
\or\qbezier(0.5,0.866)(0.446,0.12)(0.866,-0.5)
\fi
\or\ifcase#2\or\or\or
\or\qbezier(0,1.)(-0.138,0.516)(-0.5,0.866)
\or\qbezier(0,1.)(-0.26,0.45)(-0.866,0.5)
\or\qbezier(0,1.)(-0.327,0.327)(-1,0)
\or\qbezier(0,1.)(-0.311,0.179)(-0.866,-0.5)
\or\qbezier(0,1.)(-0.2,0.0536)(-0.5,-0.866)
\or\put(0, 1){\line(0, -2){2}}
\or\qbezier(0,1.)(0.2,0.0536)(0.5,-0.866)
\or\qbezier(0,1.)(0.311,0.179)(0.866,-0.5)
\fi
\or\ifcase#2\or\or\or\or
\or\qbezier(-0.5,0.866)(-0.378,0.378)(-0.866,0.5)
\or\qbezier(-0.5,0.866)(-0.45,0.26)(-1,0)
\or\qbezier(-0.5,0.866)(-0.446,0.12)(-0.866,-0.5)
\or\qbezier(-0.5,0.866)(-0.359,0)(-0.5,-0.866)
\or\qbezier(-0.5,0.866)(-0.2,-0.0536)(0,-1)
\or\put(-0.5, 0.866){\line(3, -5){1}}
\or\qbezier(-0.5,0.866)(0.146,0.146)(0.866,-0.5)
\fi
\or\ifcase#2\or\or\or\or\or
\or\qbezier(-0.866,0.5)(-0.516,0.138)(-1,0)
\or\qbezier(-0.866,0.5)(-0.52,0)(-0.866,-0.5)
\or\qbezier(-0.866,0.5)(-0.446,-0.12)(-0.5,-0.866)
\or\qbezier(-0.866,0.5)(-0.311,-0.179)(0,-1)
\or\qbezier(-0.866,0.5)(-0.146,-0.146)(0.5,-0.866)
\or\put(-0.866, 0.5){\line(5, -3){1.73}}
\fi
\or\ifcase#2\or\or\or\or\or\or
\or\qbezier(-1,0)(-0.516,-0.138)(-0.866,-0.5)
\or\qbezier(-1,0)(-0.45,-0.26)(-0.5,-0.866)
\or\qbezier(-1,0)(-0.327,-0.327)(0,-1)
\or\qbezier(-1,0)(-0.179,-0.311)(0.5,-0.866)
\or\qbezier(-1,0)(-0.0536,-0.2)(0.866,-0.5)
\fi
\or\ifcase#2\or\or\or\or\or\or\or
\or\qbezier(-0.866,-0.5)(-0.378,-0.378)(-0.5,-0.866)
\or\qbezier(-0.866,-0.5)(-0.26,-0.45)(0,-1)
\or\qbezier(-0.866,-0.5)(-0.12,-0.446)(0.5,-0.866)
\or\qbezier(-0.866,-0.5)(0,-0.359)(0.866,-0.5)
\fi
\or\ifcase#2\or\or\or\or\or\or\or\or
\or\qbezier(-0.5,-0.866)(-0.138,-0.516)(0,-1)
\or\qbezier(-0.5,-0.866)(0,-0.52)(0.5,-0.866)
\or\qbezier(-0.5,-0.866)(0.12,-0.446)(0.866,-0.5)
\fi
\or\ifcase#2\or\or\or\or\or\or\or\or\or
\or\qbezier(0,-1.)(0.138,-0.516)(0.5,-0.866)
\or\qbezier(0,-1.)(0.26,-0.45)(0.866,-0.5)
\fi
\or\ifcase#2\or\or\or\or\or\or\or\or\or\or
\or\qbezier(0.5,-0.866)(0.378,-0.378)(0.866,-0.5)
\fi\fi\fi\fi}
\def\FullChord[#1,#2]{
\Endpoint[#1]
\Endpoint[#2]
\Arc[#1]
\Arc[#2]
\Chord[#1,#2]
}
\def\EndChord[#1,#2]{
\Endpoint[#1]
\Endpoint[#2]
\Chord[#1,#2]
}
\def\Picture#1{
\begin{picture}(2,1)(-1,-0.167)
#1
\end{picture}
}
\def\DottedChordDiagram[#1,#2]{
\Picture{\DottedCircle \FullChord[#1,#2]}
}
\begin{document}
{}~
{}~
\hfill\hbox{QMUL-PH-05-14}
\break
\vskip .6cm

\centerline{{\Large \bf  Finite $N$ effects  on the collapse of
fuzzy spheres }}

\medskip

\vspace*{4.0ex}

\centerline{\large S.~McNamara,
C.~Papageorgakis,  S.~Ramgoolam and  B.~Spence ${}^{\dagger}$}

\vspace*{4.0ex}
\begin{center}
{\large Department of Physics\\
Queen Mary, University of London\\
Mile End Road\\
London E1 4NS UK\\
}
\end{center}

\vspace*{5.0ex}

\centerline{\bf Abstract} \bigskip 
\noindent 
Finite $N$ effects on the time evolution of 
 fuzzy 2-spheres moving in flat spacetime are studied using the 
non-Abelian DBI
action for $N$ $D0$-branes.
 Constancy of
the speed of light  leads to a definition of the physical radius in terms
 of symmetrised traces
of large powers of Lie algebra generators. These traces,
 which determine the dynamics at finite $N$,  have
 a surprisingly simple form.
 The energy function is given by a quotient of a free
 multi-particle system, where the dynamics of the individual 
 particles are related by a simple
 scaling of space and time. We show that exotic bounces of the kind
 seen in the $1/N$ expansion do not exist at finite $N$.
 The dependence of the time of collapse
 on $N$ is not monotonic.  The time-dependent brane acts as a source 
 for gravity which, in a region of parameter space, 
 violates the dominant energy condition. We find regimes,
 involving both slowly collapsing and rapidly collapsing 
 branes,  where higher derivative corrections to the DBI action
 can be neglected. We propose some generalised symmetrised trace
 formulae for higher dimensional fuzzy spheres and observe an  application 
 to $D$-brane charge calculations.

\thispagestyle{empty}
\vfill
\begin{flushright}
{\it ${}^{\dagger}${\{s.mcnamara, c.papageorgakis, s.ramgoolam, w.j.spence\}@qmul.ac.uk}\\
}
\end{flushright}
\eject

\renewcommand{\baselinestretch}{1.05}  


\section{Introduction}

The symmetrised trace prescription for the non-Abelian  action of
multiple  $D0$-branes
was proposed in \cite{Tseytlin} and extended
to include background RR fluxes in \cite{Myers}.
An interesting time dependent system, in which the need for an exact
prescription arises,  is a spherical bound state of $N$
 $D0$-branes 
with a spherical $D2$-brane, for finite values of $N$. This can be studied 
both from the point of view of the Abelian 
$D2$ DBI action and the non-Abelian $D0$-DBI action.  
The latter configuration also has an M-theory analogue, that of a time
dependent spherical $M2$-brane, which has been studied in the context of
matrix theory \cite{bfss,kata}.
In  \cite{rst} it was shown that the $D0$-brane 
construction, based on the fuzzy 2-sphere, agrees 
with the Abelian $D2$-construction at large $N$. 
$1/N$ corrections coming from the symmetrised trace 
 and a finite $N$ example were also studied. Here we develop further
the study of finite $N$. The need for  the non-linear 
DBI action as opposed to the Yang-Mills limit of the lower dimensional brane
was  recognised  in a spatial $D1\perp D3$ analog of the $D0-D2$ system
\cite{myersd1d3}. 

In this paper we extend the calculation of symmetrised traces 
from the spin half example of \cite{rst} to general representations
 of $SO(3)$. 
These results allow us to study in detail the finite $N$ physics of 
the time-dependent fuzzy two-sphere.
 We begin our finite $N$ analysis 
with a careful discussion on how to extract the physical 
radius from the matrices of the non-Abelian ansatz. 
The standard formula used in the Myers effect is 
$R^2 =   Tr (\Phi_i \Phi_i )/ N $. Requiring consistency with 
a constant speed of light, independent of $N$,  leads us to 
propose an equation in section 2, which agrees with the standard 
formula in large $N$ commutative limits, but disagrees in general.  
Section 3 gives  finite $N$ formulae for the energy and Lagrangian 
of the time-dependent fuzzy 2-sphere. We also give the conserved 
pressure which is relevant for the $D1\perp D3$ system. In section 4, 
we study the time of collapse as a function of $N$. In the region of large 
$N$, for fixed initial radius $R_0$,
 the time decreases as $N$ decreases. However, at some point there is a
turn-around in this trend and the time of collapse 
for spin half is actually larger than at large $N$. 
We also investigate the quantity $E^2 - p^2 $, where $E$ 
is the energy and $p$ the momentum. This quantity is of interest 
when we view the time-dependent $D$-brane as a source for spacetime fields. 
$E$ is the $T^{00}$ component of the stress tensor,
 and $p$ is the $T^{0r} $ component as we show 
by a generalisation of arguments previously used in the context 
of BFSS matrix theory.  
 For the large $N$ formulae, 
$E^2-p^2$  is always positive. At finite $N$, this can be negative, 
although  the speed of radial motion is less than the speed 
of light.  Given the relation to the stress tensor, we can interpret this 
as a violation of the dominant energy condition. 
 The other object of interest is the proper acceleration 
along the trajectory of a collapsing $D2$-brane. We find 
 analytic and numerical evidence   that there 
are regions of both large $R$ and small $R$, with small and relativistic 
velocities respectively, where the 
proper accelerations can be small. This is intriguing since 
the introduction of stringy and higher derivative effects 
in the small velocity region can be done with an adiabatic 
approximation, but it is interesting to consider approximation methods 
for the relativistic region. 

In section 5, we discuss the higher fuzzy sphere case 
\cite{gkp,clt,sphdiv,kimhigh,horam,kimura,azbag,dcp}. 
We give a general formula for $ STr ( X_i X_i )^{m} $, in general 
irreducible, representations of $SO(2k+1) $. This formula 
is motivated by some considerations surrounding $D$-brane charges 
and the ADHM construction, which are discussed in more detail in 
 \cite{simonadhm}. 
Some of the motivation is explained in Appendix A. 
This allows us a partial discussion of finite $N$ effects 
for higher fuzzy spheres. We are able to calculate the physical
 radius following the argument 
of section 2;
however, in general one needs  other 
symmetrised traces involving elements of the Lie algebra $ so (2k+1) $. 

The symmetrised trace prescription, which we study in detail in this paper,  
is known to  correctly match open string calculations up to
the first two orders in an $\alpha '$ expansion, but the correct answer
deviates from the $(\alpha ') ^3$ term onwards \cite{hata,troost, sevrin,sixgluon}.
It is possible however that for certain special symmetric 
background configurations, it may give the correct physics to all orders. 
The $D$-brane charge computation discussed in the Appendix can be
viewed as a possible 
indication in this direction. In any case, it is
important to study the corrections coming from this prescription to
all orders, in order to be able to systematically modify it, if that becomes 
necessary when the  correct non-Abelian $D$-brane action is known. 
Conversely the physics of collapsing $D$-branes can be used to constrain 
the form of the non-Abelian DBI.

\section{ Lorentz invariance and the physical radius }
We will study the collapse of a cluster of $N$ $D0$-branes in
the shape of a fuzzy $S^{2k}$, in a flat background. This configuration
is known to have a large-$N$ dual
description in terms of  spherical $D(2k)$ branes with $N$ units of
 flux. The microscopic $D0$ description  can be obtained from the
non-Abelian action  for a number of coincident
branes, proposed in \cite{Tseytlin,Myers} 
\bea\label{d1dbi}
 S_0 = -\frac{1}{g_s \ell_s} \int dt \; STr \sqrt {  - \det ( M ) }   \;,
\eea
where
\be
M =
\begin{pmatrix}
 & -1 & \lambda \partial_t \Phi_j    \\
                & - \lambda \partial_t  \Phi_i  & Q_{ij}
\end{pmatrix}  \;.
\ee
Here $a,b$ are worldvolume indices, the $\Phi$'s are worldvolume scalars,
 $\lambda = 2\pi \ell_s^{2}  $  and
\be
Q_{ij}=\delta_{ij}+i\lambda [\Phi_{i},\Phi_{j}]\;.
\ee
We will consider the time dependent ans\"atz
\be
\Phi_{i}=\hat{R}(t)  X_i \;,
\ee
The $X_i$ are matrices obeying some algebra. 
The part of the action that depends  purely on the time
derivatives and survives when $ \hat R = 0$ is 
\be 
S_{D0} = \int dt STr \sqrt { 1 - \lambda^2 (\partial_t \Phi_i)^2 }
       = \int dt STr \sqrt { 1 - \lambda^2 (\partial_t \hat R )^2 X_iX_i }\;.
\ee

 For the fuzzy $S^2$, the $X_i = \alpha_{i}$, for $i=1,2,3$, are generators
 of the irreducible  spin $n/2$  matrix
 representation of $su(2)$, with matrices of size $N= n+1$. 
 In this case the algebra is
\be\label{su2algebra}
[\alpha_{i},\alpha_{j}]=2i\epsilon_{ijk}\alpha_{k}
\ee
and following \cite{rst},  the action for $N$ $D0$-branes can be reduced to
\be\label{nonred}
S_0 = -\frac{1}{g_s \ell_s}\int dt\;STr\sqrt{1+ 4\lambda^2 {\hat R }^4 \alpha_i \alpha_i }
 \sqrt { 1 - \lambda^2 (\partial_t \hat R )^2 \alpha_i \alpha_i}\;.
\ee
If we define the physical radius using
\be\label{physrad} 
R_{phys}^2 = \lambda^2 \lim_{m \rightarrow \infty }
{ STr ( \Phi_i \Phi_i )^{m+1} \over STr ( \Phi_i \Phi_i )^{m} } 
= \lambda^2 {\hat R}^2  \lim_{m \rightarrow \infty }
{ STr ( \alpha_i \alpha_i )^{m+1} \over STr ( \alpha_i \alpha_i )^{m} },
\ee
we will find that the Lagrangian will be convergent
for speeds between $0$ and $1$. The radius
of convergence will be exactly one - this follows by applying the ratio
test to the series expansion of
\be
STr \sqrt{ 1 - \lambda^2 \dot{ \hat R}^2 \alpha_i \alpha_i }\;,
\ee
where a dot indicates differentiation with respect to time.
This leads to
\be\label{rphys}
 R_{phys}^2 =  \lambda^2 { \hat R }^2 n^2\;.
\ee
Using explicit formulae for the symmetrised traces we will also see
that, with this definition of the physical radius, the
formulae for the Lagrangian and energy will have a first singularity
at $ \dot R_{phys} =1 $.
In the large $n$ limit, the definition of physical radius
in (\ref{rphys}) agrees with   \cite{Myers}, where $R_{phys} $ 
is defined by $ R_{phys}^2  = { 1 \over N } Tr \Phi_i \Phi_i $.   
 Note that this
definition of the physical radius will also be valid for the higher
dimensional fuzzy spheres, and more generally in any matrix construction, 
where the terms in the non-Abelian DBI action depending purely on the 
velocity, are of the form 
$ \sqrt { 1 - \lambda^2 X_i X_i  ( \partial_t  \hat R )^2}$.  

In what follows, the
sums we get in expanding the square root
are conveniently written in terms of $r,s$,
 defined
by $ r^4 = 4 \lambda^2 { \hat R }^4 $ and $s^2 = \lambda^2 \dot{\hat R }^2$.
It is also useful to define
\bea\label{rphysdefs}
L^2 &=& { \lambda n \over 2 },  \nn \\
{ \hat r }^2 &=&  { R_{phys}^2 \over L^2} = { r^2  n },  \nn \\
{ \hat s }^2 &=& s^2 n^2\;.
\eea
The $ \hat r $ and $ \hat s $ variables approach the
variables  called $r,s$ in the large $n$ discussion
of \cite{rst}.


\section{ The fuzzy $S^2$ at finite $n$}\label{finite}
For the fuzzy $S^2$, the relevant algebra is that of $su(2)$, equation
\eqref{su2algebra} above. We also have the Casimir
\be
 c = \alpha_i \alpha_i = (N^2 -1 ) \nn \;,
\ee
where the last expression gives the value of the Casimir
in the $N$-dimensional representation where $N=n+1$, 
and $n$ is related to the spin $J$ by $ n = 2J $. 

We present  here the result of the
full evaluation of the symmetrised trace for odd $n$
\be\label{oddnCmn}
C(m,n) \equiv { 1 \over n+1 } STr   ( \alpha_i \alpha_i )^m
=  { 2 (2m+1) \over n + 1 } \sum_{i=1}^{ (n+1)/2 }  (2i -1 )^m \; ,
\ee
whilst for even $n$
\be\label{evennCmn}
C(m,n) \equiv
{  1 \over n +1 } STr  ( \alpha_i \alpha_i )^m =  { 2  ( 2m+1) \over n+1 } \sum_{i=1}^{ n /2 }  (2i)^m \;.
\ee
For $m=0$ the second expression doesn't have a correct analytic continuation
 and we will impose the value $STr({\alpha_i \alpha_i})^0=1$.
The expression for $C(m,1)$ was proved in \cite{rst}. A proof of
\eqref{evennCmn} for
$n=2 $ is given in Appendix B. The general formulae
given above are conjectured on the basis of
various examples, together with arguments
related to $D$-brane charges. These are given in Appendix A.
There is also a generalisation
to the case of higher dimensional fuzzy spheres,
described in section 6 and the Appendices.

We will now use the results \eqref{oddnCmn}, \eqref{evennCmn},
to obtain the  symmetrised trace
corrected energy for
a configuration of $N$ time dependent $D0$-branes blown up to
 a fuzzy $S^2$.
The reduced action (\ref{nonred}) can be expanded to give
\begin{eqnarray}
\nonumber\mathcal{L} &=& -STr\sqrt{1+ 4 \lambda^2 {\hat R }^4 \alpha_i \alpha_i }
 \sqrt { 1 - \lambda^2 \dot{\hat R}^2 \alpha_i \alpha_i} \\
&=&-STr\sqrt{1+r^{4} \alpha_i \alpha_i }\;\sqrt{1-s^{2}  \alpha_i \alpha_i }\\
&=& -STr \sum_{m=0}^{\infty}\sum_{l=0}^{\infty}s^{2m}r^{4l}(\alpha_i \alpha_i)^{m+l}\binom{1/2}{m}
\binom{1/2}{l}(-1)^{m}\;.
\end{eqnarray}
The expression for the energy then follows directly -
\begin{equation}
\mathcal{E}=-STr\sum_{m=0}^{\infty}\sum_{l=0}^{\infty}s^{2m}r^{4l}(2m-1)
(\alpha_i \alpha_i)^{m+l}
\binom{1/2}{m}\binom{1/2}{l}(-1)^{m},
\end{equation}
and after applying the symmetrised trace results
given above we get the finite-$n$  corrected energy for
any finite-dimensional irreducible representation of
spin-$\frac{n}{2}$ for the fuzzy $S^2$.

For $n=1,2$ one finds
\begin{eqnarray}\label{bscEkeq1}
 \frac{1}{2}\mathcal{E}_{n=1}(r,s)&=&\frac{1+2r^{4}-r^{4}s^{2}}
{\sqrt{1+r^4}(1-s^{2})^{3/2}}\;,\\
\frac{1}{3} \mathcal{E}_{n=2}(r,s)&=&\frac{2}{3}\;\frac{(1+8r^{4}-16r^{4}s^{2})}
{\sqrt{1+4r^{4}}(1-4s^{2})^{3/2}}+\frac{1}{3}\;.
\end{eqnarray}
We note that both of these expressions provide equations of
 motion which are solvable
by solutions of the form $\hat r=t$.
\par
For the case of general $n$,
it can be checked that the energy can be written
\be\label{energn1k1}
\mathcal{E}_n(r,s)=\sum_{l=1}^{{n+1 \over 2 } }
  \frac{2-2 ( 2l-1)^2r^4(
   (2l-1)^2s^2-2)}{\sqrt{1+ (2l-1)^2r^4}(1-(2l-1)^2s^2)^{3/2}}\;,
\ee
for $n$-odd, while for $n$ even
\be
\mathcal{E}_n(r,s) 
=  1+ 2 \sum_{l=1}^{ {n \over 2 }}
\frac{1- 8l^2 r^4( 2l^2s^2-1)}{\sqrt{1+4l^2r^4}(1- 4l^2s^2)^{3/2}}\;.
\ee
Equivalently, the closed form expression for the Lagrangian for $n$ odd
is
\be
\mathcal{L}_n(r,s)=-2\sum_{l=1}^{\frac{n+1}{2}}\frac{1-2l^2s^2+l^2r^4
(2-3l^2s^2)}{\sqrt{1+l^2r^4}\sqrt{1-l^2s^2}}\;,
\ee
whilst for $n$-even
\be
\mathcal{L}_n(r,s)=-1-2\sum_{l=1}^{\frac{n}{2}}\frac{1-2l^2s^2+l^2r^4
(2-3l^2s^2)}{\sqrt{1+l^2r^4}\sqrt{1-l^2s^2}}\;.
\ee
It is clear from these expressions that the equations of motion in the higher spin case will also admit the
$\hat r=t$ solution.
Note that, after the rescaling to physical variables (\ref{rphysdefs}),
these Lagrangians have no singularity for fixed $r$, in the region 
$ 0 \le s \le 1 $. In this sense they are consistent with a 
fixed speed of light. However, they do  not involve, for fixed $r$, the form 
$ \sqrt { dt^2 -dr^2 }$ and hence do not have an $so(1,1)$ symmetry. 
It will be interesting to see if there are generalisations of 
$so(1,1)$, possibly involving non-linear transformations 
of  $dt, dr$, which can be viewed as symmetries.     

\par
One can also get exact results for the symmetrised trace
corrected pressure of the
fuzzy-$S^{2}$ funnel configuration. The relationship between the time
dependent $D0-D2$ system and the static $D1\perp D3$ intersection was
established in \cite{largesmall}. In that paper, the large-$n$
behaviour of both systems was
described by a 
genus one Riemann surface, which is a  fixed orbit in complexified
phase space. This was done by  considering the conserved energy and
pressure and complexifying the variables $r$ and
$\partial r=s$ respectively. Conservation of the energy-momentum
tensor then yielded elliptic curves in $r,s$, involving a fixed
parameter $r_0$, which corresponded to the initial radius of the
configuration.
For our system we simply display the general
result and the first two explicit cases
\begin{eqnarray}
\mathcal{P}&=&STr\sum_{m=0}^{\infty}\sum_{l=0}^{\infty}s^{2m}r^{4l}(2m-1)(\alpha_i \alpha_i )^{m+l}
\binom{1/2}{m}\binom{1/2}{l}\\
\frac{1}{2}\mathcal{P}_{n=1}(r,s)&=&-\frac{1+2r^{4}+r^{4}s^{2}}{\sqrt{1+r^4}(1+s^{2})^{3/2}}\label{p1}\\
\frac{1}{3}\mathcal{P}_{n=2}(r,s)&=&-\frac{2}{3}\frac{(1+8r^{4}+16r^{4}s^{2})}
{\sqrt{1+4r^{4}}(1+4s^{2})^{3/2}}-\frac{1}{3}\;. \label{p2}
\end{eqnarray}
Similar results to those for the time dependent case hold for
the exact expression of the pressure for the general spin-$\frac{n}{2}$
representation. Note again that these expressions will provide
equations of motion which are solved by solutions of the form $\hat r
=1/\sigma$, where $\sigma$ is the spatial $D1$ worldvolume
coordinate. An easy way to see this is to substitute $s^2=r^4$ in
(\ref{p1}), (\ref{p2}), to find that the pressures become independent
of $r$ and $s$. Since the higher spin results for the pressure are
sums of the $n=1$ or $n=2$ cases, the argument extends.


\subsection{ Finite $N$ dynamics  as a quotient of free multi-particle dynamics }
Using the formulae above,
we can see that the fuzzy $S^2$
energy for general $n$ is determined by the energy at $n=1$. In the odd $n$ case
\be
\nn C(m,n) = \frac{2}{n+1}\ { C ( m,1 )  }
 \sum_{i_3=1}^{\frac{n+1}{2}} ( 2i_3 -1)^{2m}
 = \frac{2}{n+1}(2m+1) \sum_{i_3=1}^{\frac{n+1}{2}} ( 2i_3 -1)^{2m}\;.
\ee
Using this
form for $C(m,n)$ in the derivation of the energy, we get
\be\label{energysum}
\mathcal{E}_{n} ( r,s ) = \sum_{i_3=1}^{n+1 \over 2 }{\mathcal{E}}_{n
  =1} \left( r \sqrt
 { (2i_3 -1 )}  ~, ~ s (2i_3-1)  \right)\;.
\ee
Similarly, in the even $n$ case, we find
\be
\mathcal{E}_{n}(r,s)  = \sum_{i_3=1}^{n \over 2 }  \mathcal{E}_{n =2 }
( r \sqrt { i_3 } ~,~s(i_3)  )\;.
\ee
It is also possible to write $C(m,2)$ in terms of
$C(m,1)$ as (for $m\not= 0$)
\be
C(m,2)=\frac{2^{2m+1}}{3}C(m,1)=\frac{2^{2m+1}}{3}(2m+1)\;.
\ee
Thus we can write $\mE_{n} (r,s)$, for even $n$,
in terms of the basic  $ \mE_{n=1} (r,s) $ as
\be
\mathcal{E}_{n} ( r,s ) = 1+\sum_{i_3=1}^{n \over 2 }~
\mathcal{E}_{n =1} (r  \sqrt { ( 2i_3  )}  ~, ~ s (2i_3)  )\;.
\ee
These expressions for the energy of spin $n/2$ can be viewed as
giving the energy in terms of  a quotient of a multi-particle
system, where the individual particles are associated with the spin 
half system. 
 For example, the energy function for ${(n+1)/2}$ free particles
with dispersion relation determined by $ \mathcal{E}_{n=1} $
is $ \sum_{i} \mathcal{E}_{n=1} ( r_i , s_i ) $. By constraining
the particles by $r_i = r \sqrt{2i+1} , s_i = s(2i+1)$
 we recover precisely (\ref{energysum}).

We can now use this result to resolve a question raised 
by   \cite{rst} on the exotic bounces seen in the 
Lagrangians obtained by keeping a finite number of terms in the $1/n$ 
expansion. With the first $1/n$ correction kept, the bounce appeared 
for a class of paths involving high velocities with 
$ \gamma= { 1 \over \sqrt{ 1 - \hat s^2} }   \sim c^{1/4} $, 
near the limit of validity of the $1/n$ expansion. 
The bounce disappeared when two orders in the expansion were
kept. It was clear that whether the bounces actually happened 
or not could only be determined by finite $n$ calculations.  
These exotic bounces 
would be apparent in constant energy contour plots for $r,s$ as
a zero in the first derivative $\partial r/\partial s$.
In terms of the energies, this translates into the presence of a zero of
$\partial E/\partial s$ for constant $r$.
It is easy to show from the explicit forms of the energies that
these quantities are strictly positive for $n=1$ and $n=2$.
 Since the energy for every $n$
can be written in terms of these, we conclude that there are
no bounces for any finite $n$. This resolves the question raised in  \cite{rst} 
about the fate at finite $n$ of these bounces.

We note that
the large-$n$ limit of the formula for the  energy provides
 us with a consistency check.  In the large
 $n$-limit the sums above become integrals. For the odd-$n$ case
 (even-$n$ can be treated in a similar fashion),
 define $ x = { 2 i_3-1 \over n }
 \sim { 2i_3 \over n } $. Then the sum in \eqref{energysum}
 goes over to the integral
\be
{ n \over 2 } \int_0^1
 dx  { 2 - 2x^2n^2r^4 ( x^2n^2s^2- 2 ) \over \sqrt{ 1 + x^2n^2r^4 }
( 1 - x^2s^2 )^{3/2} }
=  { n  \sqrt { 1 + r^4 n^2 } \over \sqrt{ 1  - s^2 n^2 } }\;.
\ee
By switching to the $\hat r, \hat s$ parameters
the energy can be written as
$ n \sqrt { 1 + {\hat r}^4  } \over \sqrt{ 1  - {\hat s}^2  }$.
This matches
exactly the large $n$ limit used in \cite{rst}.


\section{ Physical properties of the finite $N$ solutions }
\subsection{ Special limits where finite $n$ and
 large $n$ formulae agree }

In the above we compared the finite $n$ formula
with the  large $n$ limit.
 Here we consider the comparison between
the fixed $n$ formula and the large $n$ one in some other limits.
 On physical grounds
we expect some agreement. The $D0-D2$ system at large $r$ and small
velocity $s$
is expected to be correctly described by the $D2$ equations. These
coincide with the large $n$ limit of the $D0$.
In the $D1\perp D3$ system, the large $r$ limit with large imaginary $ s $
is also described by the $D3$.

Such an argument should extend to the finite-$n$ case. In
 \cite{largesmall}, these systems were simply described by a genus
 one Riemann surface. However, in this case
the energy functions are more complicated and the resulting Riemann
surfaces are of higher genus. We still expect the region 
 of the finite $n$ curve,
with large
$r$ and small, real $s$, to agree with the
same limit of the large $n$ curve. We also expect the region
of large $r$ and large imaginary $s$ to agree with large $n$.

For concreteness consider odd $n$.
Indeed for large $r$, small $s$, (\ref{energn1k1}) gives
\be\label{4.1}
\sum_{l} 4 (2l-1) r^2  \sim  n r^2\;,
\ee
which agrees with $  { \hat r }^2 $.
In this limit, both the genus one curve and the high
genus finite $n$ curves degenerate to a pair of points.
Now consider large $r$ and large imaginary $s$.
This is the right regime since the $D1\perp D3$ system is described
by $ r \sim { 1 \over \sigma } $ which means that $r$
is large at small $ \sigma $, where $ { dr \over d \sigma } = i s $ is large.
For $ s = i S $
\be
\mathcal{P} \sim - n r^2 /S\;,
\ee
which agrees with the same limit of the large $n$ curve.
In this limit, both the large $n$ genus one curve
and the finite $n$ curves of large genus degenerate to
a genus zero curve.

The agreement in (\ref{4.1}) between the $D0$ and $D2$ pictures
is a stringy phenomenon. It follows from the fact that
there is really one system, a bound state of $D0$ and $D2$ branes.
A boundary conformal field theory would have boundary conditions that
encode the presence of both the $D0$ and $D2$.
In
the large $N$ limit, the  equations of motion
 coming from the $D0$-effective action
agree with the $D2$-effective action description at all $R$.
This is because at large $N$ it is possible to specify a DBI-scaling
where the regime of validity of both the $D0$ and $D2$ effective
actions extends for all $R$. This follows because the DBI
scaling has $ \ell_s \rightarrow 0 $ \cite{prt}.
Indeed it is easy to see that the effective open string metric
discussed in \cite{prt} has the property
that $ \ell_s^2 G^{-1} = { \ell_s^2 R^2 \over R^4 + L^4 } $ goes
to zero when $ N \rightarrow \infty $ with $L= \ell_s \sqrt {\pi N
},R $ fixed.
This factor  $ \ell_s^2 G^{-1} $  controls higher
derivative corrections for the open string degrees of freedom.
At finite $N$, we can keep $\ell_s^2 G^{-1} $ small, either
when $ R \ll L $ or  $ R \gg L $. Therefore, there are two regimes
where the stringy description reduces to an effective field theory,
where higher derivatives can be neglected.
 The agreement holds for specified regions of
$R$ as well as $s$, because the requirement
$\ell_s^2 G^{-1} \ll 1 $ is not
the only condition needed to ensure that higher derivatives can be neglected.
We also require that the proper acceleration is small. 
At large $R$, the magnetic flux density is small (as well as the higher 
derivatives being small)  and the 
$D2$-brane without non-commutativity is a good description. This is why the 
finite $N$ equations derived from the $D0$-brane effective field theory 
agree with the  Abelian $D2$-picture.  
 For small $R$, small $s$, we can also neglect higher derivatives.
This is the region where the $D0$-Yang-Mills description is
valid, or equivalently
a strongly non-commutative $D2$-picture.

\subsection{Finite $N$  effects : Time of collapse, proper accelerations and 
 violation of the dominant energy condition}

We will consider the time of collapse as a function of
$n$ using  the definition of the physical
radius given in section (2).
In order to facilitate comparison with the
large $n$ system, we will be using $\hat r, \hat s$ variables.
To begin with, consider the dimensionless  acceleration, which can be
expressed as
\be
- \hat s \frac{\partial_{\hat s} \mathcal E |_{\hat
  r}}{\partial_{\hat r} \mathcal E
|_{\hat s}}\;,
\ee
with $\gamma=1/\sqrt{(1-\hat s^2)}$.
As the sphere starts collapsing from $\hat r=\hat r_0$ down to $\hat r=0$,
the speed changes from $\hat s=0$ to a value less than  $ \hat s= 1 $.
It is easy to see that the acceleration does not change sign
in this region. Using the basic energy $\hat {\mE}=\mathcal E/N$ from
(\ref{bscEkeq1}), we can write
\bea
{ \partial  \hat {\mE}_{n=1}(\hat r,\hat s)  \over \partial \hat s }
&=& \hat s { ( 3 (1+\hat r^4) + \hat r^4 ( 1-\hat s^2 ) ) \over { \sqrt{
( 1 +\hat r^4 ) }       ( 1-\hat s^2)^{ 5 \over 2}  } }, \nn \\[6pt]
{ \partial  \hat {\mE}_{n=1}(\hat r,\hat s)  \over \partial \hat r }
&=& { 2\hat r^3 \over { (1+\hat r^4)^{3 \over 2} ( 1-\hat s^2)^{3 \over 2} }}
    \bigg( ( 1+\hat r^4) + (1-\hat s^2)(2+\hat r^4)  \bigg)\;.
\eea
Neither of the  partial derivatives change sign in the range
$\hat s=0$ to $1$. Hence the speed $\hat s$  increases monotonically.
The same result is true for $n >1 $, since the energy functions
for all these cases can be written as a sum of the energies
at $n=1$.

In the $n=1$ case $, \hat r = r $, $ \hat s = s $. For fixed $r_0$ the
speed  at  $ r=0$ is given by
\be
 ( 1 -   s^2|_{n=1} )
 =   {  ( 1  +   r_0^4 )^{1 \over 3 } \over ( 1 + 2 r_0^4 )^{2 \over 3 }  }.
\ee
Comparing this with the large $n$ formula
\be
( 1 -s^2|_{n=\infty} ) =  ( 1 + r_0^4 )^{-1},
\ee
it is easy to see that
\be
 \left ( { ( 1- s^2 )|_{n= \infty} \over ( 1 -s^2 )|_{n= 1 } } \right )^3
= { ( 1 + 2r_0^4 )^2 \over  ( 1 +  r_0^4 )^4 } < 1,
\ee
which establishes that the speed at $r=0$ is
larger for $n= \infty$.

We can strengthen this result to show that
the speed of collapse at all $ r < r_0 $ is smaller
for $n=1$ than at $ n = \infty$.  For any $r < r_0$ we evaluate
this energy function with the speed of collapse
evaluated at $ s^2 = { r_0^4 - r^4 \over  r_0^4 +1 } $, which
is the speed at the same $r$ in the large $n$ problem. Let us
define $ F( r , r_0 ) = \hat { \cE }_{n=1}  \left( r , s =
 \sqrt  { r_0^4 - r^4 \over  r_0^4 +1 } \right) $. We compare this
with $ \hat { \cE }_{n=1} (r, s ) $ for $s$ appropriate for the
$n=1$ problem, which is just $ { 1 + 2r_0^4 \over \sqrt { 1 + r_0^4 } } \equiv
 G ( r_0 )  $
by conservation of energy.
We now use the fact, established
above,  that $ { \partial \hat { \cE }_{n=1}  \over \partial s } $
is positive for any real $r$. This means that we can show
$s|_{n=1} < \sqrt  { r_0^4 - r^4 \over  r_0^4 +1 }$ by showing that
$F(r,r_0) > G ( r_0 )$. A short calculation gives
\be
F ( r , r_0 ) - G ( r_0 )
= { r_0^4 \over { \sqrt { 1 + r_0^4 } ( 1 + r^4 )} } ( r_0^4 - r^4 ).
\ee
It is clear that we have the desired inequality, showing
that, at each $r$, the speed $s$ in the $n=1$ problem
is smaller than the speed in the $n= \infty $ system.
Hence the time of collapse is larger at  $n=1$.
In the $n=2$ case, we find that an exactly equivalent treatment
proves again that the collapse is slower than at large $n$. However,
this trend is not a general feature for all $n$. In the leading 
large-$N$ limit, the time of collapse is given by the formula
\be
{ T\over L } =\int dr \frac{\sqrt{1+r_0^4}}{\sqrt{r_0^4-r^4}} = 
{ K ( { 1\over \sqrt{2}} ) \over \sqrt{2} } { \sqrt { R^4 + L^4 } \over R }
\ee
For fixed $\ell_s$, $ L$ decreases with decreasing $N$ and as a result 
$T$ decreases. 
When we include the first $1/N$ correction
the time of collapse is  \cite{rst}
\be
{ T \over L }  =\int dr\left[ \frac{\sqrt{1+r_0^4}}{\sqrt{r_0^4-r^4}}+\frac{r_0^8}{6 N^2
  (1+r_0^4)^{3/2}\sqrt{r_0^4-r^4}}-\frac{r_0^4(1+3(1+r_0^4))}{6 N^2
  (1+r^4)\sqrt{1+r_0^4}\sqrt{r_0^4-r^4}}\right]\;.
\ee
By performing numerical integration of the above
for several values of the parameter $r_0$ and some large but finite
values of $N$, we see that the time of collapse is
smaller for the $1/N$ corrected case. This means that, in the region of 
large $N$ the time of collapse decreases as $N$ decreases, with both the 
leading large $N$ formula and the $1/N$ correction 
being consistent with this trend.  
However, as we saw above the time of collapse at $n=1$ 
and $n=2$ are larger than at  $n=\infty$. This means that 
there are one or more turning points 
in the time of collapse as a function of $n$.

The deceleration effect that arises in the comparison 
of $n=1 $ and $n=2$  with large $n$ may have applications in cosmology. 
Deceleration mechanisms coming from DBI actions
have been studied in the context of bulk causality
in AdS/CFT \cite{ts1,kl}
and applied in the problem of satisfying
slow roll conditions in stringy inflation \cite{ts2}. Here we see
that the finite  $n$ effects result in a further deceleration
in the region of small $n$.

We turn to the  proper acceleration which is important in checking
the validity of our action. Since the DBI action is 
valid when higher derivatives are small,  it is natural to demand that the
proper acceleration,  should be small (see for example \cite{ts1}).
 The condition  is
$ \gamma^3 \ell_s  \partial_t^2  R_{phys} \ll 1 $.
 In terms of the dimensionless
variables it is $ \gamma^3  ( \partial_\tau^2 \hat r ) \ll  \sqrt N $.
If we want a trajectory with initial radius $  r_0$ such that the proper
acceleration always remains less than one through the collapse,
then there is an upper bound on $r_0$. This upper bound goes to infinity
as $ N \rightarrow \infty $. We are already constrained by the
condition that the spatial derivatives are small, to lie within the
small or large $r$ region, for finite-$N$. For small $r_0$ we are in the
matrix theory limit and things are well behaved. For large $r_0$ and
$r$-large, the acceleration is under control, $\alpha\sim1/r$ 
 and the velocity will be close to zero. Interestingly, there will also be
a valid large $r_0$, relativistic regime. Consider for example
the $n=1$ case. The proper acceleration can be written as
\be
\alpha  = - \frac{2r^3}{1+r^4}\frac{-3+2 s^2 +r^4
  (s^2-2)}{\sqrt{1-s^2}
(r^4(s^2-4)-3)}\;.
\ee
For $s\sim 1$ and small $r$, this becomes
\be
\alpha \simeq-\frac{2 r^3}{3\sqrt{1-s^2}}
\ee
and $\sqrt{1-s^2}$ can be found from the energy at the same limits,
in which (\ref{bscEkeq1}) becomes
\be
\sqrt{1-s^2}\simeq\frac{1}{(2 r_0^2)^{1/3}}\;.
\ee
Therefore, we can identify a region where  the proper acceleration is
small by restricting it to be  of order $ 1/r_0$ for example
\be
\alpha \simeq \frac{2 r^3}{3}(2 r_0^2)^{1/3}\sim \frac{1}{r_0}\;.
\ee
This means that in regions where $r\sim r_0^{-5/9}$, we will have a
relativistic limit described by the DBI, where stringy corrections 
can be neglected. This result also holds in the
large-$N$ limit. It will be interesting to develop a perturbative approximation
which systematically includes stringy effects away from  this region. 

Another quantity of interest is 
the effective mass squared  $ E^2 - p^2  $
, where
$p=\partial \mathcal L/\partial s$ is the radial conjugate
momentum. It becomes negative for sufficiently large velocities. 
This includes the above regime of relativistic speeds 
and small radii. 
 It is straightforward to see that if our collapsing
configuration is considered as a source for spacetime gravity, this implies a
violation of the dominant energy condition. In the context of the BFSS
matrix model, it has been shown that for an action containing a
background spacetime $G_{IJ}=\eta_{IJ}+h_{IJ}$,
 in the linearised approximation, linear couplings in
the fluctuation $h_{0I}$ correspond to momentum in the $X^I$ direction 
\cite{tayvan}. 
The same argument can be developed here for the non-Abelian DBI. 
We couple a  small fluctuation $h_{0r}$, which in classical geometry 
we can write as $h_{0i} = h_{0r}x_i $ for the unit sphere. 
We replace $x_i $ by $ \alpha_i/ n $.   
 The action for $D0$-branes \cite{Tseytlin,Myers} is generalised from 
 (\ref{d1dbi}) by replacing $ \dot R  $ in 
 $\lambda \partial_t \Phi_i = \lambda ( \dot{ \hat R } ) \alpha_i 
=  { \dot  R \over n  } \alpha_i $ 
with $   ( \dot R + h_{0r})   $.    It is then clear that 
the variation with respect to $\dot R $, which gives $p$, is the same as the 
variation with respect to $h_{0r} $, which gives $T^{0r} $.   
Hence, the dominant energy condition will
 be violated, since  $ E < |p|$  is equivalent to 
  $T^{00}<T^{0r}$.  The  violation of this condition 
 by stringy $D$-brane matter can have 
 profound consequences. For a discussion of possible consequences
 in cosmology see \cite{tach}.
 In  this context, it is noteworthy that the violation  can occur near a 
 region of zero radius, which could be relevant 
to a near-big-bang region in a braneworld scenario.


\subsection{Distance to blow-up in $D1\perp D3$ }

Comparisons between the finite  and large $N$
results can be made in the spatial case using the
conserved pressure. The arguments are similar to what we used 
for the time of collapse using the energy functions. 
 Consider the case $ n=1$, and let
$\hat P = P /N $.     First calculate
the derivative of the  pressure -
\be
 { \partial { \hat { P } }   \over \partial s } =
  { s ( 4r^4 + r^4 s^2 +3 ) \over \sqrt {1 + r^4} ( 1 -s^2)^{5/2} }.
\ee
This is clearly always positive.
Now evaluate
\be
 \hat P \left( r  , s =  {\sqrt { r^4 - r_0^4 } \over \sqrt { 1 + r_0^4 } }  \right)
= - { ( 1 +r_0^4)^{1/2} \over 1 + r^4 } ( 1 + r_0^4 + r^4 ).
\ee

This should be compared with
$ \hat P ( r,s) $, evaluated for the value of $s$ which solves the
$n=1$ equation of motion, which by conservation of pressure
is $ - { ( 1 + 2r_0^4) \over \sqrt { 1 + r_0^4 } } $.

Take the difference to find
\be
\hat P \left( r  , s = { \sqrt { r^4 - r_0^4 } \over \sqrt { 1 + r_0^4 } } \right)
+ { ( 1 + 2r_0^4) \over \sqrt { 1 + r_0^4 } }
 = { r_0^4 ( r^4 - r_0^4 ) \over \sqrt { 1 + r_0^4 } ( 1 + r^4 ) }.
\ee

Thus at fixed $r_0$ and $r$,
 $ \hat P_{n=1}$, when  evaluated for the value of
$s$ which solves the large $n$ problem,  is larger than
when it is evaluated for the value of $s$
which solves the $ n = \infty $ problem.
Since $\hat P$ increases monotonically with $s $ for fixed $r$,
this shows that for fixed $r_0$, and any $r$,  $s$ is always larger
in the large $N$ problem. Since
$\Sigma = \int dr /s $, this means the distance to blow-up
is smaller for $ n = \infty $.
Hence for fixed $r_0$, the distance to blow-up is larger
at $n=1$.

\section{Towards a generalisation to higher even-dimensional fuzzy-spheres}

For  generalisations to higher dimensional brane systems, and to
higher dimensional fuzzy spheres \cite{sphdiv,horam,kimura,azbag}, it
is of interest to derive an extension of the
expressions for the symmetrised traces given above.
In the general case,
we define $N(k,n)$ to be the dimension of the  irreducible representation
of $SO(2k+1)$ with
Dynkin label $(\frac{n}{2},\frac{n}{2},\dots,\frac{n}{2})$ which
contains $k$ entries. We then take  $C(m,k,n)$ to be the action of
the symmetrised trace on $m$ pairs of matrices $X_i$, where $i=1,\ldots,
2k+1$
\begin{equation}
C(m,k,n)=\frac{1}{N(k,n)}STr \big(X_iX_i)^m.
\end{equation}
Finding an expression for $C(m,k,n)$ is non-trivial. Investigations
based upon intuition from the ADHM construction lead us to conjecture that
for  $n$ odd
\begin{equation}\label{guessodd}
C(m,k,n)=\frac{2^k
  \prod_{i_1=1}^{k}(2m-1+2i_1)}{(k-1)!\prod_{i_2=1}^{2k-1}(n+i_2)}
\sum_{i_3=1}^{\frac{n+1}{2}} \Big[
  \prod_{i_4=1}^{k-1}\Big(\Big(\frac{n}{2}+i_4\Big)^2
-\Big(i_3-\frac{1}{2}\Big)^2\Big)(2i_3-1)^{2m} \Big]\;,
\end{equation}
while for $n$ even\footnote{For $m=0$ the value $STr({X_i X_i})^0=1$
  is once again imposed.}
\begin{equation}\label{guesseven}
C(m,k,n)=\frac{2^k
  \prod_{i_1=1}^{k}(2m-1+2i_1)}{(k-1)!\prod_{i_2=1}^{2k-1}(n+i_2)}
\sum_{i_3=1}^{\frac{n}{2}} \Big[
  \prod_{i_4=1}^{k-1}\Big(\Big(\frac{n}{2}+i_4\Big)^2-i_3^2\Big)
(2i_3)^{2m} \Big]\;.
\end{equation}
We give the arguments leading to the expressions above in Appendix A.

For higher even spheres there will be extra complications at
finite-$n$. Consider the case of the fuzzy $S^4$ for
concreteness.
The evaluation of the higher dimensional determinant in the
corresponding non-Abelian brane action will
give expressions with higher products of $\partial_t
\Phi_i$ and $\Phi_{ij} \equiv [\Phi_i, \Phi_j]$ 
\bea\label{s4}
S &=&-T_0\int dt \,STr\left\{
1+\lambda^2(\partial_t\Phi_i)^2+2\lambda^2\Phi_{ij}\Phi_{ji}
+2\lambda^4(\Phi_{ij}\Phi_{ji})^2-4\lambda^4\Phi_{ij}\Phi_{jk}
\Phi_{kl}\Phi_{li}+\phantom{1\over4}\right.\nonumber\\
&&\!\!\left.
\phantom{1}+2\lambda^4(\partial_t\Phi_i)^2\Phi_{jk}
\Phi_{kj}-4\lambda^4\partial_t\Phi_i\Phi_{ij}\Phi_{jk}
\partial_t\Phi_{k}+{\lambda^6\over4}(\epsilon_{ijklm}
\partial_t\Phi_i\Phi_{jk}\Phi_{lm})^2\right\}^{1/2}\;.
\eea
 The ans\"atz for the transverse scalars will still be
\be
\nn \Phi_i=\hat R(t) X_i\;,
\ee
where now $i=1,\ldots, 5$ and the $X^i$'s  are given by the  action of
$SO(5)$ gamma matrices on the totally symmetric $n$-fold tensor
product of the basic spinor.  After expanding the square root,
the symmetrisation procedure should take place over all the $X_i$'s and
 $[X_i,X_j]$'s. However, the commutators of commutators $[[X,X],[X,X]]$
will give a  nontrivial contribution, as opposed to what happens in the large-$n$
limit where they are sub-leading and are taken to be zero. Therefore, in order to uncover the full answer for the
finite-$n$ fuzzy $S^4$ it is not enough to just know the result
of $STr(X_iX_i)^m$ - we need to know the full
$STr\left((X\;X)^{m_1}([X,X][X,X])\right)^{m_2}$ with all
possible contractions among the above. It would be clearly interesting
to have the full answer for the  fuzzy $S^4$. A similar story
will apply for the higher even-dimensional fuzzy spheres.

Note, however, that for $ \hat R =0$ in (\ref{s4}) all the commutator terms $\Phi_{ij}$
will vanish, since they scale like ${ \hat R } ^2$. This 
reduces the symmetrisation procedure to the one involving $X_iX_i$ and
yields  only one sum for the energy. The same will
hold for any even-dimensional $S^{2k}$, resulting in the following
general expression
\bea
\nn \mathcal{E}_{n,k}(0,s)&=& - STr\sum_{m=0}^{\infty}(-1)^m s^{2m}(2m-1)
(X_iX_i)^{m} \binom{1/2}{m}\\
&=& -N(k,n)\sum_{m=0}^{\infty}(-1)^m s^{2m}(2m-1)C(m,k,n)\binom{1/2}{m}\;.
\eea
Using (\ref{guessodd}), notice that
in the odd $n$ case
\be
C(m,k,n) = \prod_{i_2=1}^{2k-1} { ( 1+i_2) \over (n+i_2) }
           { C ( m,k,1 )  } \sum_{i_3}^{\frac{n+1}{2}}
            { f_{odd}  ( i_3, k , n ) \over f_{odd} (1,k,1) } ( 2i_3 -1)^{2m}\;.
\ee
The factor $f_{odd}$ is
\be
f_{odd}  ( i_3, k , n ) = \prod_{i_4=1}^{k-1}\Big(\Big(\frac{n}{2}+i_4\Big)^2
-\Big(i_3-\frac{1}{2}\Big)^2\Big)\;.
\ee

Inserting this form for $C(m,k,n)$ in terms of $ C ( m,k,1 ) $
we see that
\be\label{oddenergyasum}
\mathcal{E}_{n,k } ( 0,s ) = N(n,k)
  \prod_{i_2=1}^{2k-1} { ( 1+i_2) \over (n+i_2) }
\sum_{i_3=1}^{n+1 \over 2 }~
{ f_{odd} (i_3,k, n) \over f_{odd} (1, k, 1 ) }   ~
 \hat{\mathcal{E}}_{n =1 ,k } (0, ~ s (2i_3-1)  )
\ee

Similarly we derive, in the even $n$ case, that
\be\label{evenenergyasum}
\mathcal{E}_{n,k }(0,s)  = N(n,k)
 \prod_{i_2=1}^{2k-1} { ( 2+i_2) \over (n +i_2) }
 \sum_{i_3=1}^{n \over 2 } { f_{even} (i_3,k,n) \over f_{even} ( 1,k,2 ) }
\hat{ \mathcal{E}}_{n =2 ,k } ( 0,~s(i_3)  )
\ee
where
\be
f_{even}(i_3,k,n) =  \prod_{i_4=1}^{k-1}\Big( \Big(\frac{n}{2}+i_4\Big)^2 - i_3^2\Big)
\ee
 and $\hat{\mathcal{E}}$ is the energy density, i.e. the energy
 divided a factor of $N(n,k)$.

It is also possible to write
$C(m,k,2) $ in terms of  $C(m,k,1) $
\bea
C(m,k,2) &&= 2^{2m}
 C(m,k,1 ) \prod_{i_4=1}^{k-1} { i_4(i_4+2) \over i_4(i_4+1) }
 \prod_{i_2=1}^{2k-1} {  ( i_2 +1) \over (i_2+2) }  \nn \\
    &&= 2^{2m}
 C(m,k,1 ) { f_{even} ( 1,k,2) \over f_{odd} ( 1,k,1 ) }
 \prod_{i_2=1}^{2k-1} {  ( i_2 +1) \over (i_2+2) }\; , 
\eea
which is valid for all values of $m \neq 0 $. 

It turns out to be possible to give
explicit forms for
the energy for the $n=1$ and $n=2$ case. Since the definition of the
physical radius in section 2 is also valid for higher dimensional fuzzy
spheres, we can express  the results in terms of the rescaled
variables $\hat r$ and $\hat s$
\bea
\nn \hat
    {\mathcal{E}}_{n=1,k}(0,\hat s)&=&\frac{1}{(1-\hat s^2)^{\frac{2k+1}{2}}}\\
 \hat
    {\mathcal{E}}_{n=2,k}(0,\hat s)&=&\frac{1}{(1-\hat s^2)^{\frac{2k+1}{2}}}\frac{(k+1)}{(2k+1)}\;.
\eea
When plugged into (\ref{oddenergyasum}), (\ref{evenenergyasum}) 
the above results provide a closed form for the energy at $\hat r=0$, for any $n$ and
any $k$.

\section{Summary and Outlook  }

We have given a detailed study of the finite $N$ effects
for the time dependent $D0-D2$ fuzzy sphere system and the related
$D1\perp D3$ funnel. This involved calculating symmetrised traces
 of $SO(3)$ generators. The formulae have a surprising
 simplicity.

 The energy function  $ E(r,s) $ in the large $N$ limit
looks like a relativistic particle with position dependent mass.
This relativistic nature is modified at finite $N$.
Nevertheless our results are consistent with a fixed
relativistic upper speed limit. This is guaranteed by an
appropriate definition of the physical radius which
relies on the properties of symmetrised traces of large
numbers of generators.
We showed that the exotic bounces found in the large $N$
expansion in \cite{rst} do not occur. It was previously clear that
these exotic bounces happened near the regime where the
$1/N$ expansion was breaking down. The presence or absence of these
could only be settled by a finite $N$ treatment, which we have provided
in this paper.
We also compared the time of collapse of the finite $N$ system
with that of the large $N$ system and found a finite $N$
deceleration effect for the first small values of $N$.
The modified $E(r,s)$ relation allows us to define an
effective squared mass which depends on both $r,s$.
For certain regions in $(r,s)$ space, it can be negative.
When the $D0-D2$ system is viewed as a source for gravity,
a negative sign of this effective mass squared indicates
that the brane acts as a gravitational source which violates the dominant
energy condition.

We extended some of our discussion to
the case of higher even fuzzy  spheres with $SO(2k+1)$
symmetry. The results
for symmetrised traces that we obtain can be used
in a proposed calculation of charges in the $D1\perp D(2k+1)$
system. They also provide further illustrations
of how the correct definition of physical radius
using symmetrised traces of large powers of Lie algebra
generators gives consistency with a constant speed of light.
A more complete discussion of the finite $N$ effects
for the higher fuzzy spheres could start from these results.
Generalisations of the finite $N$ considerations
to fuzzy spheres in more general backgrounds \cite{js}  
will be interesting to consider, with a view to possible applications 
in cosmology.

\bigskip

{\bf Acknowledgements}:
We would like to thank Jan de Boer, Simon Nickerson and 
 John Ward  for useful discussions.
 The work of SR is supported by
a PPARC Advanced Fellowship. SM and CP would like to acknowledge a
QMUL Research Studentship. This work was in
part supported by the EC Marie Curie Research Training Network
MRTN-CT-2004-512194 {\it Superstrings}, and by the PPARC Research Grant
PPA/G/O/2002/00477  {\it M Theory, String Theory and
Duality}.

\newpage
\begin{appendix}
\section{ General formula for the symmetrised trace }

As in section 5, we define $N(k,n)$ to be the dimension of the irreducible representation of
$SO(2k+1)$ with Dynkin label $(\frac{n}{2},\frac{n}{2},\dots,\frac{n}{2})$.
 These are the usual fuzzy sphere
representations \cite{sphdiv,kimhigh} (for example, for k=1 the
$X_i$ are the elements of the Lie algebra of SU(2) in the irreducible
representation with spin $\frac{n}{2}$). Then
\begin{equation}\label{eq:bign}
N(k,n) = \prod_{1 \le i < j \le k} \frac{n+2k-(i+j)+1}{2k-(i+j)+1}
\prod_{l=1}^k \frac{n+2k-2l+1}{2k-2l+1}\;.
\end{equation}

The symmetrised trace is defined to be the normalised sum over permutations
of the matrices
\begin{equation}
STr\big(X_{i_1}\dots X_{i_p})= \frac{1}{p!}
\sum_{\sigma \in S_{p}} Tr (X_{\sigma(1)}\dots X_{\sigma(p)})\;.
\end{equation}

We have given earlier a conjecture for the symmetrised trace
of $m$ powers of the
quadratic Casimir $X_iX_i=c\; \mathbb{I}_{n\times n}$, where $c=n(n+2k)$. This is,
for all $m,k$ and $n$ even
\begin{equation}\label{eq:guesseven}
\frac{1}{N(k,n)}STr\big(X_iX_i)^m=\frac{2^k \prod_{i_1=1}^{k}
(2m-1+2i_1)}{(k-1)!\prod_{i_2=1}^{2k-1}(n+i_2)}\sum_{i_3=1}^{\frac{n}{2}}
\Big[ \prod_{i_4=1}^{k-1}\Big(\Big(\frac{n}{2}+i_4\Big)^2-i_3^2\Big)(2i_3)^{2m}
 \Big]   \;,
\end{equation}
where for $k=1$ the product over $i_4=1,\ldots,k-1$ is just defined to be
equal to $1$. Similarly for all $m,k$ and for $n$ odd we have proposed that
\begin{equation}\label{eq:guessodd}
\frac{1}{N(k,n)}STr\big(X_iX_i)^m=\frac{2^k \prod_{i_1=1}^{k}
(2m-1+2i_1)}
{(k-1)!\prod_{i_2=1}^{2k-1}(n+i_2)}\sum_{i_3=1}^{\frac{n+1}{2}} \Big[
\prod_{i_4=1}^{k-1}\Big(\Big(\frac{n}{2}+i_4\Big)^2-\Big(i_3-\frac{1}{2}
\Big)^2\Big)(2i_3-1)^{2m} \Big].
\end{equation}
The argument leading to these conjectures follows.
Firstly, it is possible to view the fuzzy sphere
matrices $X_i$ as the transverse
 coordinates of the world-volume theory of a stack of $D1$-branes expanding
  into a stack of $D(2k+1)$-branes \cite{myersd1d3,myersd1d5,robertd1d7}. There is also a dual realisation of this
   system in which the $D1$-branes appear as a monopole in the world-volume
    theory of the $D(2k+1)$-branes.
The ADHM construction can be used to construct the monopole dual to the fuzzy
 sphere transverse coordinates. If one takes the $N(k,n)$-dimensional fuzzy
 sphere matrices representing a stack of $N(k,n)$ $D1$-branes as ADHM data
 for a monopole, then one naturally constructs a monopole defined on a stack
  of $N(k-1,n+1)$ $D(2k+1)$-branes.
We have also done the calculation which shows that the
charge for the  monopole just constructed gives precisely 
  $N(k,n)$, which provides a consistency check 
 (this calculation extends some results in \cite{Kimura:2003ab} 
   and will appear in \cite{simonadhm}).

It is also possible to calculate the number of $D(2k+1)$ 
branes from the fuzzy sphere ansatz for the transverse coordinates, by looking at the RR coupling
on the $D$-string worldvolume.
In this case one does not get $N(k-1,n+1)$ as one would expect, but
instead the quantity
\begin{equation}\label{eq:wrongnumber}
N(k-1,n+1)\frac{\prod_{i=1}^{2k-1}(n+i)}
{c^{k-\frac{1}{2}}},
\end{equation}
where $c=n(n+2k)$. This number of branes does agree with
$N(k-1,n+1)$ for the first two orders in the large $n$ expansion
\begin{equation}
\textrm{Number of $D(2k+1)$-branes}=N(k-1,n+1)\Big(1+O\Big(\frac{1}
{n^2}\Big)\Big)\;.
\end{equation}

Now consider this RR charge calculation more carefully.
First take the $k\!=\!1$ case. Based on the ADHM
construction we expect the number of $D3$-branes
to be $N(0,n+1)\!=\!1$. However, equation
(\ref{eq:wrongnumber}) suggests that the charge calculation
gives for $k=1$ the answer
\begin{equation}
\frac{n+1}{c^{\frac{1}{2}}}\;.
\end{equation}
Suppose that the numerator in
the above is correct, but that the denominator is correct only at
large $n$ and that it receives corrections at lower order to make the
number of $D3$-branes exactly one. Then these corrections
need to satisfy
\begin{equation}
1=(n+1)(c^{-\frac{1}{2}}+x_1c^{-\frac{3}{2}}+x_2c^{-\frac{5}{2}}+....).
\end{equation}
It is easy to show that we need $x_1=-\frac{1}{2}$ and $x_2=\frac{3}{8}$,
 by Taylor expanding and using that $c=n(n+2)$ for $k=1$. Therefore,
 we would like to have a group theoretic justification for the series
\begin{equation}\label{eq:series}
c^{-\frac{1}{2}}-\frac{1}{2}c^{-\frac{3}{2}}+\frac{3}{8}c^{-\frac{5}{2}}+...\;.
\end{equation}
There exists a formula for the first three terms in the large $n$ expansion of the
$k=1$ symmetrised trace operator \cite{rst}, namely
\begin{equation}\label{eq:rst}
\frac{1}{N(1,n)}STr(X_iX_i)^m=c^m-\frac{2}{3}m(m-1)
c^{m-1}+\frac{2}{45}m(m-1)(m-2)(7m-1)c^{m-2}+...\;.
\end{equation}
Now, if we make the choice $m\!=\!-\frac{1}{2}$ in (\ref{eq:rst}) we get precisely
 (\ref{eq:series}).
 However, this suggests that if this choice is
 correct, then we should have an all orders prediction for the action of
 the symmetrised trace operator. Thus, for $k\!=\!1$ we predict that
\begin{equation}\label{eq:conjecture1}
\frac{1}{N(1,n)}STr(X_iX_i)^m\Big|_{m=-\frac{1}{2}} \simeq \frac{1}{(n+1)}\;,
\end{equation}
where for future reference we consider the left hand side to be equal
to the symmetrised trace in a large-$n$ series expansion, as appeared in \cite{rst}.

Checking the conjecture (\ref{eq:conjecture1}) beyond the first
 three terms in a straightforward fashion, by techniques
 similar to those employed in \cite{rst}, proves difficult.
This involves either adding up a large number
 of chord diagrams, or complicated combinatorics if one uses the highest
 weight method.

  An alternative approach involves first writing down the conjecture based on
  brane counting for general $k$, since the methods of \cite{rst}
turn out to generalise  from the $k=1$ to the general $k$ case.
  The conjecture for general $k$, based on the
    brane counting, follow immediately from (\ref{eq:wrongnumber})
\begin{equation}\label{eq:conjecture}
\frac{1}{N(k,n)}STr(X_iX_i)^m\Big|_{m=-k+\frac{1}{2}}
\simeq \prod_{i=1}^{2k-1}\frac{1}{(n+i)}\;.
\end{equation}
Note that the right hand side of this equation
 appears in the factor outside the sum in  (\ref{eq:guesseven}) and
 (\ref{eq:guessodd}).
 Notice also that the above expression concerns the large $n$
 expansion of the symmetrised trace considered at $m=-k+\frac{1}{2}$.

One can repeat the $k=1$ calculation of \cite{rst} for
general $k$, to check the first three terms of this conjecture. A sketch
of this calculation follows before displaying the full results.
 First we calculate
$\frac{1}{N(k,n)}STr(X_iX_i)^m$ for $m\!=\!2,3,4$.
Then we find the first three terms in the symmetrised trace, 
 large $n$ expansion using these results.
 Finally we can check that the conjecture (\ref{eq:conjecture}) is
  true for the first three terms in
 the symmetrised trace
  large $n$ expansion, for general $k$ as well as for $k=1$.
We then proceed to calculate the fourth term in the expansion,
for general $k$. To do this we need to calculate
$\frac{1}{N(k,n)}STr(X_iX_i)^m$ for $m=5,6$.
 We then show that the fourth term in the large $n$ expansion of the
 symmetrised trace agrees precisely with (\ref{eq:conjecture}).

In the following, we use the notation
of \cite{rst}, with each trace of a string of $2m$ $X_i$
matrices arising here being represented
by a chord diagram with $m$ chords.
This provides a convenient way to represent equivalent strings of matrices.

For the calculation of $STr(X_iX_i)^2$
there are three different strings of the four $X_i$ matrices and two different
 chord diagrams. Two of the three strings correspond to the same chord
 diagram. In the following, the first column contains a
 fraction which is the multiplicity of
 the chord diagram in the list of strings divided by
 the total number of strings. The second column
 contains a picture of the chord diagram preceded
  by an example of a string in the equivalence class
  defined by this chord diagram. The evaluation of the chord
  diagram is the final entry.
\begin{equation*}
\begin{array}{llll}
\frac{2}{3}&1122&=&\Picture{\FullCircle\EndChord[4,8]\EndChord[2,10]}=c^2,\\[20pt]
\frac{1}{3}&1212&=&\Picture{\FullCircle\EndChord[3,9]\EndChord[0,6]}=(c-4k)
\Picture{\FullCircle\EndChord[3,9]}=c(c-4k)\;.\\[20pt]
\end{array}
\end{equation*}
Using this, one finds immediately that
\begin{equation}\label{eq:power2}
\frac{1}{N(k,n)}STr(X_iX_i)^2=\quad c^2-\frac{4}{3}kc\;.
\end{equation}
For $m\!=\!3$ there are 15 different strings of $X_i$ matrices
and five different chord diagrams, which evaluate as follows
\begin{equation*}
\begin{array}{llll}
\frac{2}{15} & 112233 & =
& \Picture{\FullCircle\EndChord[4,6]\EndChord[2,0]\EndChord[8,10]}=c^3, \\[20pt]
\frac{6}{15} & 112323 &
= & \Picture{\FullCircle\EndChord[4,8]\EndChord[2,11]
\EndChord[1,10]}=c \,\Picture{\FullCircle\EndChord[3,9]\EndChord[0,6]}=c^2(c-4k), \\[20pt]
\frac{3}{15} & 112332
& = & \Picture{\FullCircle\EndChord[4,8]\EndChord[3,9]\EndChord[2,10]}=c^3,\\[20pt]
\frac{3}{15} & 121323
 & = & \Picture{\FullCircle\EndChord[0,6]\EndChord[2,10]
 \EndChord[4,8]}=(c-4k)\Picture{\FullCircle
 \EndChord[0,6]\EndChord[3,9]}=c(c-4k)^2, \\[20pt]
\frac{1}{15} & 123123
& = & \Picture{\FullCircle\EndChord[3,9]\EndChord[1,7]\EndChord[11,5]}=c^3-12kc^2+16k(k+1)c \;.\\[20pt]
\end{array}
\end{equation*}
Thus we find that
\begin{equation}\label{eq:power3}
\frac{1}{N(k,n)}STr(X_iX_i)^3=c^3-4kc^2+\frac{16}{15}k(4k+1)c\;.
\end{equation}
For $m\!=\!4$ there are 105 different strings and 18 different chord diagrams.
We omit the details for simplicity.
The final result is that
\begin{equation}\label{eq:power4}
\frac{1}{N(k,n)}STr(X_iX_i)^4=c^4 -8kc^3
+\frac{16}{5}k(7k+2)c^2-\frac{64}{105}k(34k^2+24k+5)c.
\end{equation}
For $m\!=\!5$ there are 945 different strings of $X_i$ matrices,
    and 105 different chord diagrams\footnote{We acknowledge the assistance of
Simon Nickerson, for writing a computer programme used here. Maple files
for these calculations are available from the authors.}. The result is
\begin{eqnarray}\label{eq:power5}
\frac{1}{N(k,n)}STr(X_iX_i)^5&=&\phantom{+}c^5
-\frac{40}{3}kc^4
+\frac{16}{3}(13k+4)kc^3
-\frac{64}{63}(158k^2+126k+31)kc^2\nonumber\\
&&+\frac{256}{945}(496k^3+672k^2+344k+63)kc.
\end{eqnarray}
For $m\!=\!6$ there are 10395 different strings of $X_i$ matrices,
and 902 different chord diagrams, and we find that
\begin{eqnarray}\label{eq:power6}
\frac{1}{N(k,n)}STr(X_iX_i)^6&=&\phantom{+}c^6
-20kc^5+\frac{16}{3}(31k+10)kc^4
-\frac{64}{63}(677k^2+582k+157)kc^3\nonumber\\
&&+\frac{256}{315}(1726k^3+2616k^2+1541k+336)kc^2\nonumber\\
\nn &&-\frac{1024}{10395}(11056k^4+24256k^3+22046k^2+9476k+1575)kc\;.\\
\end{eqnarray}

Now we calculate the first four terms in the large $n$ expansion of
$STr(X_iX_i)^m$. Suppose that the coefficient of
$c^{m-l}$ term in $STr(X_iX_i)^m$ is a polynomial in
$m$ of order $2l$. Then we have the following ans\"atz: The known factors of these
 polynomials come from the fact that the series has to terminate
  so that there are never negative powers of $c$ for $m\!=\!1,2,3,...$ Then
\begin{eqnarray*}
\frac{1}{N(k,n)}STr(X_iX_i)^m &=&\phantom{+}c^m+y_1(k)m(m-1) c^{m-1}\\
&&+\Big(y_2(k)m+y_3(k)\Big) m(m-1)(m-2) c^{m-2}\\
&&+\Big(y_4(k)m^2+y_5(k)m+y_6(k)\Big) m(m-1)(m-2)(m-3) c^{m-3}\\
&&+O(c^{m-4})\;.
\end{eqnarray*}
We now find the unknown functions $y_1(k),y_2(k) \ldots y_6(k)$
using the results of $STr(X_iX_i)^m$ for $m=2,3,4,5,6$
calculated above. We find that
\begin{eqnarray}\label{thisone}
y_1(k)&=&-\frac{2}{3}k, \quad  y_2(k) = \frac{2}{45}(5k+2)k,  \nonumber \\
y_3(k) &= &\frac{2}{45}(k-2)k, \quad y_4(k) =  \frac{1}{2835}(-140k^2-168k-64)k,  \nonumber \\
\nn y_5(k) &=&  \frac{1}{2835} (-84k^2+216k+192)k, \quad   y_6(k) =\frac{1}{2835}(128k^2+96k-104)k\;.\\
\end{eqnarray}

Now we are able to
 provide a check of the conjecture (\ref{eq:conjecture}).
 First we express the right-hand 
 side of (\ref{eq:conjecture}) as a function of $c$ as
\begin{equation}\label{thatone}
\prod_{l=1}^{2k-1}\frac{1}{n+l} = \frac{1}{\surd(k^2+c)}\prod_{l=1}^{k-1}\frac{1}{c+2kl-l^2}
=c^{-k+\frac{1}{2}}\sum_{j=0}^{\infty}\frac{b_j}{c^j}\;,
\end{equation}
where
\begin{eqnarray}\label{eq:answerrhs}
b_1&=&-\frac{2}{3}k\Big(k-\frac{1}{2}\Big)\Big(k+\frac{1}{2}\Big), \nonumber\\
b_2&=&\frac{1}{45}(10k^2-3k+2)k\Big(k-\frac{1}{2}\Big)\Big(k+\frac{1}{2}\Big)
\Big(k+\frac{3}{2}\Big), \nonumber\\
b_3&=&\frac{1}{2835}(-24+34k-61k^2+56k^3-140k^4)\times\nonumber\\
&&\quad   k\Big(k-\frac{1}{2}\Big)\Big(k+\frac{1}{2}\Big)\Big(k+\frac{3}
{2}\Big)\Big(k+\frac{5}{2}\Big)\;.
\end{eqnarray}
Now consider the left-hand side of (\ref{eq:conjecture}) involving
the large $n$ expansion of
$STr(X_iX_i)^m$, which we calculated above, but now we set $m=-k+\frac{1}{2}$.
Expanding in inverse powers of $c$, we find that this becomes
\begin{equation*}
\frac{1}{N(k,n)}STr(X_iX_i)^m\Big|_{m=-k+1/2}
=c^{-k+\frac{1}{2}}\sum_{j=0}^{\infty}\frac{b_j}{c^j}\;,
\end{equation*}
with precisely the coefficients $b_i$ given in
\eqref{eq:answerrhs}.
Given the extensive and non-trivial calculations required to obtain
these results, we believe that there is strong evidence for the truth of (\ref{eq:conjecture}).

For $k=1$ the guess of the exact answer for $n$ even is
\begin{equation}\label{eq:guesseven1}
\frac{1}{N(1,n)}STr(X_iX_i)^m=
\frac{2(2m+1)}{n+1}\sum_{i=1}^{\frac{n}{2}} (2i)^{2m}\;.
\end{equation}
It is easy to show that (\ref{eq:guesseven1}) agrees with the
 first four orders in the large $n$ expansion
(\ref{thisone}) for $k=1$. If we set $m=-\frac{1}{2}$ in this we get zero,
because of the $(2m+1)$ factor. This might
appear to contradict  (\ref{eq:conjecture1}),
but it is easy to show, using a large $n$ expansion, that if
 (\ref{eq:guesseven1}) is true then
(\ref{eq:conjecture1}) holds to all orders. To calculate the large $n$ 
expansion of this sum we can use the Euler-Maclaurin
formula. This approximates the sum by an integral, plus an infinite series of
corrections involving the Bernoulli numbers $B_{2p}$
\begin{eqnarray}
\sum_{i=1}^{n}f(i)&\simeq &\int_{0}^{n+1}f(x)\textrm{d}x+\frac{1}{2}[f(n+1)-f(0)]\nonumber\\
&&+\sum_{p=1}^{\infty}\frac{B_{2p}}{(2p)!}[f^{(2p-1)}(n+1)-f^{(2p-1)}(0)]\;.
\end{eqnarray}
We see from this calculation that for $k=1$ the value $m=-\frac{1}{2}$
is very special. It is the only value of $m$ for which the higher order terms 
in the Euler-MacLaurin  large $n$  approximation
 of the sum in (\ref{eq:guesseven1}) are zero.

\section{ Finite $n$ results on symmetrised traces from the highest weight method }

Results on finite $n$ symmetrised traces
can be obtained by generalising the highest weight method
of \cite{rst}.
For the $SO(3)$ representations used in fuzzy 2-spheres we have
\be
{1 \over 2}STr_{J=1/2}  ( \alpha_i \alpha_i )^m  = ( 2m +1 )   \;,
\label{sphaf}
\ee
where the $1/2$ comes from dividing with the dimension of the spin-$1/2$
representation. A similar factor will appear in all of the results below.
The above result was derived in \cite{rst}.
For the spin one case, we will obtain
\be
{1\over 3}STr_{ J=1 }  ( \alpha_i \alpha_i )^m  = { 2^{2m+1} ( 2m +1 )\over 3 }    \;.
\label{spinone}
\ee
These results can be generalised to representations
of $SO(2l+1)$ relevant for higher fuzzy spheres.
The construction of higher dimensional fuzzy spheres
uses irreducible representations of highest weight $ ( {n \over 2 } , \cdots ,
 { n \over 2 } ) $, as we have noted. 
 For the minimal representation with $ n =1 $ we have
\be
{1\over D_{n=1}}STr_{n=1} ( X_i X_i ) =  { ( 2l  + 2m -1) !! \over ( 2m -1) !! ( 2l -1) !! }
\label{minrepl} \;.
\ee
 Notice the interesting symmetry
under the exchange of $l$ and $m$.
For the next-to-minimal irreducible representation with $n=2$ we obtain:
\be
{1\over D_{n=2}}STr_{n=2} ( X_i X_i ) = 2^{2m}  ( l+1 )   { ( 2l  + 2m -1) !!
\over ( 2m -1) !! ( 2l + 1) !! }
\label{nxtminrep} \;.
\ee
This is a generalisation of the spin $1$ case to higher orthogonal groups.
It agrees with the formulae in section A of the appendix, 
with $ l \rightarrow k $.

\subsection{Review of spin half for $SO(3)$ }
We will begin by recalling some facts about the
derivation  of the $n=1$ case in \cite{rst}.
The commutation relations can be expressed in terms
of $ \alpha_{3} , \alpha_{\pm } $
\bea
\alpha_{\pm } &=& { 1\over \sqrt{2} } ( \alpha_1 \pm i \alpha_2 ), \nn \\
\left[ \alpha_3 , \alpha_{ \pm } \right]  &=&  2 \alpha_{\pm}, \nn \\
 \left[ \alpha_+ , \alpha_- \right] &=&  2 \alpha_3, \nn \\
 c &=& \alpha_{+} \alpha_{-} +\alpha_{-} \alpha_{+} + \alpha_{3}^2. \nn \\
\eea
With these normalisations, the eigenvalues of $ \alpha_3 $ in the spin half
representation are $ \pm 1 $ and $ \alpha_+ \alpha_- $ is $1$ on the highest
weight state.

It is useful to define a quantity $ \tilde C ( p , q ) $
which depends on two natural numbers $p,q$ and
counts the number of ways of separating $p$ identical objects
into $ q $ parts
\be
\tilde C ( p , q ) = { ( p + q -1 ) !\over p ! ( q -1) ! } \;.
\ee
We begin by a review of the spin half case, establishing
a counting which will be used again in  more complicated
cases below. This relies on a sum
\bea
2^k \sum_{\hat i_{2k}}^{2n- 2k }
\cdots \sum_{\hat i_2 =  0 }^{\hat i_3 } \sum_{\hat i_1 = 0 }^{\hat i_2}
 (-1)^{ \hat i_1 + \hat i_2 + \cdots
\hat i_{2k} } =  2^k { n! \over (n-k ) ! k ! }   \nnm
\label{rstsum}\; .
\eea
Recall that this sum was obtained by evaluating
a sequence of generators of $SO(3)$ consisting of
$k$ pairs $ \alpha_{-} \alpha_{+} $ and with powers
of $\alpha_{3}$ between these pairs -
\be
\alpha_3^{2J_{2k+1}}  \alpha_{+} \alpha_3^{2J_k} \alpha_-  \cdots
      \alpha_{-}   \alpha_3^{J_3}
      \alpha_{+}   \alpha_3^{J_2}  \alpha_- \alpha_{3}^{J_1} \; .
\ee
We can move the powers of $\alpha_3 $ to the left
to get factors $ ( \alpha_3 -2 )^{J_2 + J_4 + \cdots J_{2k} }$.
Moving the $\alpha_3$ with powers $ J_1 , J_3 .. $
gives $ \alpha_3^{J_1 + J_3 + \cdots } $. The $k$ powers of
 $ \alpha_- \alpha_+ $ gives $2^k $.  The above sum
can be rewritten
\bea
2^k \sum_{J_{2k+1 } =0 }^{2m- 2k }
\cdots \sum_{ J_2 =  0 }^{ 2m-2k - J_3 + .. J_{2k+1}  }
\sum_{ J_1 = 0 }^{2m-2k - J_2 ... J_{2k+1}  }
 (-1)^{ J_2 + J_4 \cdots
 J_{2k}  } =  2^k { n! \over (n-k ) ! k ! }
\label{rstsum1}\; .
\eea
This includes a sum over $ J_e = J_2 + J_4 + \ldots +J_{2k} $.
The summand does not depend on the individual $ J_2, J_4,\ldots $
only on the sum $J_e$ which ranges from $0$ to $2m-2k$.
The sum over $ J_2, J_4, \ldots $ is the combinatoric
factor, introduced above,  which is the number of ways of splitting $ J_e $
identical objects into $k$ parts, i.e. $ \tilde C ( J_e , k ) $.
The remaining $2m - 2k - J_e $ powers of $\alpha_3$ are distributed
in $ k +1 $ slots in $ \tC ( 2m - 2k - J_e , k+1) $ ways. Hence the
sum (\ref{rstsum1}) can be written more simply as
\be
2^k \sum_{J_e=0 }^{2m - 2k } (-1)^{J_e} \tilde C ( J_e , k ) \tilde C
(2 m - 2k -  J_e , k +1 ) = 2^k  { m! \over (m-k ) ! k ! }
\label{simprst} \; .
\ee
Then there is a sum over $k$ from $0$ to $m$, with weight
\be\label{cfac}
C(k,m) = { 2^k k! (2m-2k)! m! \over (m-k)! (2m)!   }
\ee
which gives the final result $2m+1$  \cite{rst}.
Similar sums arise in the proofs below.
In some cases, closed formulae for the sums are
obtained experimentally.

\subsection{ Derivation of symmetrised trace for
 minimal $SO(2l+1)$ representation }

The Casimir of interest here is
\be
X_{\mu} X_{ \mu } = X_{2l +1}^2 +
\sum_{i=1}^{l} \left( X_{-}^{(i)}X_{+}^{(i)} +
X_{+}^{(i)}X_{-}^{(i)} \right) \;.
\ee
The patterns are similar to those above, with
$ \alpha_3$ replaced by $ X_{2l+1}$, and noting that here there
are $l$ ``colours'' of $ \alpha_{\pm } $  which are
$ X_{\pm}^{(l)}$. All the states in the fundamental spinor
are obtained by acting on a vacuum which is annihilated by
$l$ species of fermions.
Generally we might expect patterns
\be
... X_{2l+1}^{i_1} X_{-}^{(j_1)} X_{2l+1}^{i_2} X_{+}^{(j_2)} ...
\ee
In evaluating these, we can commute all the $ X_{2l+1} $ to the left.
This results in shifts which do not depend on the value of $j$.
It is easy to see that whenever $X_{+}^{(1)}$ is followed by
$ X_{+}^{(1)} $ we get zero because of the fermionic construction of
the gamma matrices. $X_{+}^{(1)} $ cannot also be followed by $ X_{+}^{(2)}$
because $ X_{+}^{(1)} X_{+}^{(2)} + X_{+}^{(2)} X_{+}^{(1)} =0 $.
So the pairs have to take the form $  X_{-}^{(j)}  X_{+}^{(j)} $ for fixed
$j$.
The sum we have to evaluate is
\bea
&&\sum_{k=0}^{m } \sum_{J_e = 0 }^{2m - 2k } ( -1)^{J_e} \tC( J_e , k )
\tC ( 2m - 2k - J_e , k )  \tC ( k , l ) 2^{k } C ( k , m  ) \nnm \\
&& = \sum_{k=0}^m  { m \choose k } 2^k C ( k , m  ) \tC ( k , l )\\
&& =  { ( 2l  + 2m -1) !! \over ( 2m -1) !! ( 2l -1) !! } \nnm \;.
\eea
The factors $\tC( J_e , k )$ and  $ \tC ( 2m - 2k - J_e , k )$
have the same origin as in the spin half case.
The factor  (\ref{cfac}) is now generalised to an $l$-colour version
\be
 C ( k_1 , k_2 ... k_l ; m ) = 2^k { ( 2m - 2k ) ! \over ( 2m ) ! } { m ! \over
( m - k ) ! } k_1 ! k_2 ! ... k_l !
\ee
This has to be summed over $ k_1 ,.. ,k_l$.
For fixed $k = k_1 + \cdots + k_l $ we have
\be
\sum_{ k_1 .. k_l }   C ( k_{1} .. k_l  , m  ) { k! \over k_1! .. k_l! }
=  \sum_{ k_1 .. k_l }   C ( k , m  )   =  C ( k , m  )  \tC ( k , l )\;.
\ee
The combinatoric factor ${ k! \over k_1! .. k_l! }$ in the second line
above comes from the different ways of distributing the
$k_1 .. k_l $ pairs of $(-+)$ operators in the $k$ positions
along the line of operators. The subsequent sum amounts
to calculating the number of ways of separating $k$ objects
into $l$ parts which is given by  $\tC ( k , l )$.
The $C(k,m) $ is familiar from (\ref{cfac}).
This sum can be done for various values of $k,m$ and gives
agreement with (\ref{minrepl}).

\subsection{Derivation of spin one symmetrised trace for $SO(3)$}
For the spin one case more patterns will arise.
After an $ \alpha_-$ acts on the highest weight, we get
a state with $ \alpha_3 =0 $ so that we have, for any positive $r$
\be
\alpha_3^r \alpha_{-} |J=1, \alpha_3 = 2 > = 0\;,\qquad \forall\quad r > 0 \;.
\ee
Hence any $\alpha_- $  can be followed immediately
by $ \alpha_{+}$. These neutral pairs of $ ( \alpha_+\alpha_{-})$
can be separated by powers of  $\alpha_{3}$.
 Alternatively
an $ \alpha_-$ can be followed immediately by $ \alpha_-$.
The effect of $ \alpha_-^2$ is to change the highest weight
state to a lowest weight state. In describing the patterns
we have written the ``vacuum changing operator'' on the second line,
with the first line containing only neutral pairs separated by $ \alpha_3$'s.
Let there be $J_1$ neutral pairs  in this first line and $L_1$ powers of
$\alpha_3$ distributed between them.
After the change of vacuum, we can have a
sequence of $(\alpha_-\alpha_+)$   separated by powers
of $ \alpha_3$. Let there be a total of $J_2$ neutral pairs and
$L_2$  $\alpha_3$'s in the second line.
 At the beginning of the third line we have another
vacuum changing operator $ \alpha_+^2  $ which takes us back to the highest
weight state. In the third line, we have $J_3$ neutral pairs and
$L_3$ powers of $\alpha_3$. The equation below describes a general
pattern with $p$ pairs of vacuum changing operators. The total
number of neutral pairs is $2p + J$ where
 $ J = J_1 + J_2 + \cdots + J_{2p+1}$.
The general pattern of operators acting on the vacuum is
\bea
{\#}~ (\alpha_+\alpha_-) ~{\#} ~ (\alpha_+\alpha_{-})~{\#}~  \cdots
 {\#}  ~ (\alpha_+\alpha_{-} ) ~ {\#}  &&  | J=1  , \alpha_3 =2  > \nnm\\
{\#}~  (\alpha_-\alpha_+)~  {\#} ~  ( \alpha_-\alpha_{+} )~  {\#}  \cdots
 {\#}  ~ ( \alpha_-\alpha_{+} )~  {\#} \alpha_-^2 ~~~~&&  \nnm\\
{\#} ~ (\alpha_+\alpha_-)~ {\#}~   ( \alpha_+\alpha_{-}) ~  {\#}  \cdots
 {\#}  ~ ( \alpha_+\alpha_{-}) ~  {\#} ~ \alpha_+^2 ~~~~~~&&  \nnm\\
\vdots \hskip.3in && \nnm \\
{\#} ~ (\alpha_-\alpha_+) ~  {\#}  ~  ( \alpha_-\alpha_{+}) ~  {\#} ~ \cdots
 ~ {\#} ~   ( \alpha_-\alpha_{+}) ~  {\#} \alpha_-^2 ~~~~~~~~&&  \nnm\\
{\#} ~ (\alpha_+\alpha_-)~  {\#}~   ( \alpha_+\alpha_{-} )~  {\#}  \cdots
 {\#} ~   ( \alpha_+\alpha_{-} )  {\#}~  \alpha_+^2 ~~~~~~~~~~~~&&
\label{spin1gpt} \;,
\eea
where in the above the first line of operators acts on the state 
$| J=1  , \alpha_3 =2  >$ first, then the second line acts, and so on.
The  symbols $ {\#}$ represent powers of $\alpha_{3}$.
We define $J_e = J_2 + J_4 +\cdots J_{2p} $ which is the
total number of $(-+)$ pairs on the even lines above.
There is a combinatoric factor $ \tilde C ( J_e, p ) $
for distributing $J_e$ among the $p$ entries, and a similar
$ \tilde C ( J - J_e, p +1  ) $ for the odd lines. The
$ L_e = L_2 + L_4 + \cdots + L_{2p} $ copies of  $\alpha_3$ can sit
in $ ( J_2+1) + (J_4+1 ) + \cdots +  ( J_{2p} +1 ) $ positions
which gives a factor of $ \tilde C ( L_e , J_e +p ) $.
The $  L_1 + L_3 + \cdots + L_{2p+1}$ can sit
in $ ( J_1 +1 ) + (J_3+1 ) + \cdots + ( J_{2p+1} +1 ) = J-J_e +p+1 $
positions, giving a factor
$ \tilde C ( 2m - 2J - 4p - L_e , J - J_e + p +1 ) $.
There is finally a factor $ C ( 2p+ J , m ) $ defined in
(\ref{cfac}) which arises from the number of different
ways the permutations of $2m$ indices   can be specialised to yield
a fixed pattern of $ \alpha_+ , \alpha_- , \alpha_3 $
\bea
 \sum_{p=0}^{ [m/2]} \sum_{J = 0 }^{ m - 2p } \sum_{J_e=0}^{J}
\sum_{L_e = 0 }^{2m - 4p - 2J }  &&
\tC ( J_e, p ) ~~ \tC ( J - J_e , p+1 ) \nnm 
\;( -1)^{L_e} ~~ \tC ( L_e , J_e + p )\times \\ 
&&\tC ( 2m - 2J - 4p - L_e , J - J_e + p +1 )\times \nnm \\
&&2^{ 2m - 2J - 4 p } ~~  Q( 1,1 )^{J - J_e } ~~  Q( 2,1)^{J_e} ~~
Q(2,2)^p \; C ( 2p + J , m ) \nnm \;.
\eea
By doing the sums (using Maple for example) for various values of
$m$ we find  $ { 2^{2m+1} ( 2m+1 )  \over 3 } $.
The factors  $Q(i,j)$, denoted in \cite{rst} by $N(i,j)$,
arise from evaluating the $ \alpha_- , \alpha_+ $ on the
highest weight.

\subsection{ Derivation of next-to-minimal representation for $SO(2l+1)$}

\vskip.2in
The $n=2$, general $l$   patterns are again similar to the
$n=2, l=1$ case except that the $\alpha_{-}, \alpha_+ $ are replaced by
coloured objects of $l$ colours, i.e. the $ X_{\pm}^{(j)} $.
We also have the simple replacement of $\alpha_3$   by $X_{2l+1}$.

We define linear combinations of the gamma matrices which are
 simply related to a set of $l$ fermionic oscillators:
 $ \Gamma_{+}^{(i)} =  { 1\over \sqrt{2} } ( \Gamma_{2i-1}  +
 i \Gamma_{2i} ) =   \sqrt{2}a_{i}^{\dagger}   $ and
 $  \Gamma_{-}^{(i)} =  { 1\over \sqrt{2} } ( \Gamma_{2i-1}  -
 i \Gamma_{2i} ) =  \sqrt{2} a_{i}  $. As usual $ X_i $ are expressed
 as operators acting on an $n$-fold tensor product, and
\be
X_{\pm}^{(i)} = \sum_r \rho_r ( \Gamma_{ \pm }^{(i)} ) \;.
\ee
Some useful facts are
\bea
X_{2l+1}^r X_+ |0 > &=& 0, \qquad  X_{2l+1}^r X_+^2 |0> \;\;\; =\;\;\; (-2)^r X_+^2 |0>,\nnm\\
X_{2l+1}^r X_{-}X_{+} |0> &=& X_{-}X_{+} X_{2l+1}^{r} |0>\;\;\; =\;\;\;
(2)^r  X_{-}X_{+} |0>,  \nnm\\
X_{-}X_{+} X_{2l+1}^r X_{+}^2 |0 > &=& 0, \qquad  Y_{+} X_{+}^2 + X_{+}^2Y_{+} |0> \;\;\; =\;\;\;0,   \nnm \\
X_{+}Y_{+}X_{+} |0>  &= &0, \qquad   X_{-}Y_{+}X_{+} |0>  \;\;\; =\;\;\; 0,     \nnm \\
X_+X_{-} X_{+}^2 |0> &=& Q(2,1) X_+^2 |0>, \qquad   X_+X_{-} X_{+}Y_{+} |0 > \;\;\; =\;\;\; Q(2,1) X_+Y_{+} |0>,         \nnm \\
X_{-}^2X_{+}^2 |0 > &=& Q(2,2) |0 >,\qquad  Y_{-}X_{-}X_{+}Y_{+}|0 > \;\;\; =\;\;\; Q(2,2) |0 > \nnm\;.
\label{usfids}
\eea
It is significant that the
same $Q(2,1), Q(2,2) $ factors appear in the different places
in the above equation.
In the above $X_{+}$ stands for any of the $l$
$X_{+}^{(i)} $'s. Any equation containing
$X_{\pm}$ and $Y_{\pm}$  stands for any pair
$X_{\pm}^{(i)}$ and $X_{\pm}^{(j)} $ for $i,j$ distinct
integers from $1$ to $l$.

The general pattern is similar to (\ref{spin1gpt})
with the only difference that the $ ( \alpha_{-}\alpha_{+} ) $
on the first line is replaced by any one $  ( X^{(i)}_- X^{(i)}_{+} ) $
for $ i = 1,\ldots, l $. The positive vacuum changing operators can be
$ (X^{(i)}_{+}X_{+}^{(j)}) $, where $i,j$ can be identical or different.
For every such choice the allowed neutral pairs following them
are  $ X^{(j)}_{+}X_{-}^{(i)}  $ and the dual vacuum changing operator
is  $ (X^{(j)}_{-}X_{-}^{(i)}) $.

The summation we have to do is:

\bea
\sum_{p=0}^{ [m/2]}
 \sum_{J = 0 }^{ m - 2p }  \sum_{L_e = 0 }^{2m - 4p - 2J } \sum_{J_e=0}^{J}
 && \bigg( C ( 2p + J , m )\tC ( 2p + J , l )
\tC ( J_e, p ) ~~ \tC ( J - J_e , p+1 ) \times \nnm  \\
&& ( -1)^{L_e} ~~ \tC ( L_e , J_e + p )  
\tC ( 2m - 2J - 4p - L_e , J - J_e + p +1 ) \times \nnm  \\[3pt]
&&   2^{ 2m - 2J - 4 p } ~~  Q( 1,1 )^{J - J_e } ~~  Q( 2,1)^{J_e} ~~
Q(2,2)^p\bigg) \nnm \;.
\eea

The $Q$-factors can be easily evaluated on the highest weight and
then inserted into the above
\be
Q(1,1)=4\;,\qquad Q(2,1)=4\;\qquad Q(2,2)=16  \;.
\ee
By computing this  for several values of
$m,l$, we obtain (\ref{nxtminrep}). Note that both the $l=1$ and the
general $l$ case will yield the correct value for $m=0$, which is 1.

\end{appendix}


\end{document}